\documentclass{article}

\usepackage[preprint]{corl_2021} %

\usepackage{times}
\usepackage{epsfig}
\usepackage{graphicx}
\usepackage{amsmath}
\usepackage{amssymb}
\usepackage{arydshln}
\usepackage{soul}

\usepackage{algorithm}
\usepackage{algorithmic}
\usepackage{array}
\usepackage{eqparbox}

\usepackage{etoolbox}  %
\makeatletter
\patchcmd{\algorithmic}{\addtolength{\ALC@tlm}{\leftmargin} }{\addtolength{\ALC@tlm}{\leftmargin}}{}{}
\makeatother

\usepackage{float}
\usepackage[utf8]{inputenc} %
\usepackage[T1]{fontenc}    %
\usepackage{url}            %
\usepackage{booktabs}       %
\usepackage{amsfonts}       %
\usepackage{nicefrac}       %
\usepackage{microtype}      %
\usepackage{color}
\usepackage[table]{xcolor}
\usepackage{soul}
\usepackage{tabularx}
\usepackage{nicematrix}
\usepackage{arydshln}
\usepackage{multirow}
\usepackage{microtype}
\usepackage{wrapfig}
\usepackage[font=small]{caption}

\usepackage{natbib}

\newcommand{\mypar}[1]{\textbf{#1}}

\newcommand{\ci}[1]{\tiny{\textcolor{gray}{~($\pm #1$)}}}
\newcommand{\ourrow}{\rowcolor{gray!10}}
\newcommand{\ourcell}{\cellcolor{gray!10}}

\def\audiodataset{Quiet Campus}

\def\audioum{static}
\def\audiotd{motion}
\def\capaudioum{Static}
\def\capaudiotd{Motion}

\usepackage{titlesec} 

\titlespacing\section{0pt}{1pt plus 0.3pt minus 0.3pt}{0pt plus 0.3pt minus 0.3pt}
\titlespacing\subsection{0pt}{0.3pt plus 0.2pt minus 0.2pt}{0pt plus 0.2pt minus 0.2pt}
\titlespacing\subsubsection{0pt}{0.2pt plus 0.1pt minus 0.1pt}{0pt plus 0.1pt minus 0.1pt}

\makeatletter
\newcommand{\printfnsymbol}[1]{%
  \textsuperscript{\@fnsymbol{#1}}%
}
\makeatother

\title{Structure from Silence: Learning \\ Scene Structure from Ambient Sound}

\author{
  Ziyang Chen\textsuperscript{*}, Xixi Hu\textsuperscript{*}, Andrew Owens\\
  University of Michigan \\
  {\small \url{https://ificl.github.io/structure-from-silence}}
}

\begin{document}

\renewcommand{\footnoterule}{\vfill\kern -3pt \hrule width 0.4\columnwidth \kern 0.5pt}

\maketitle
\begin{center}
    \vspace{-2.0em}
\includegraphics[width=\textwidth]{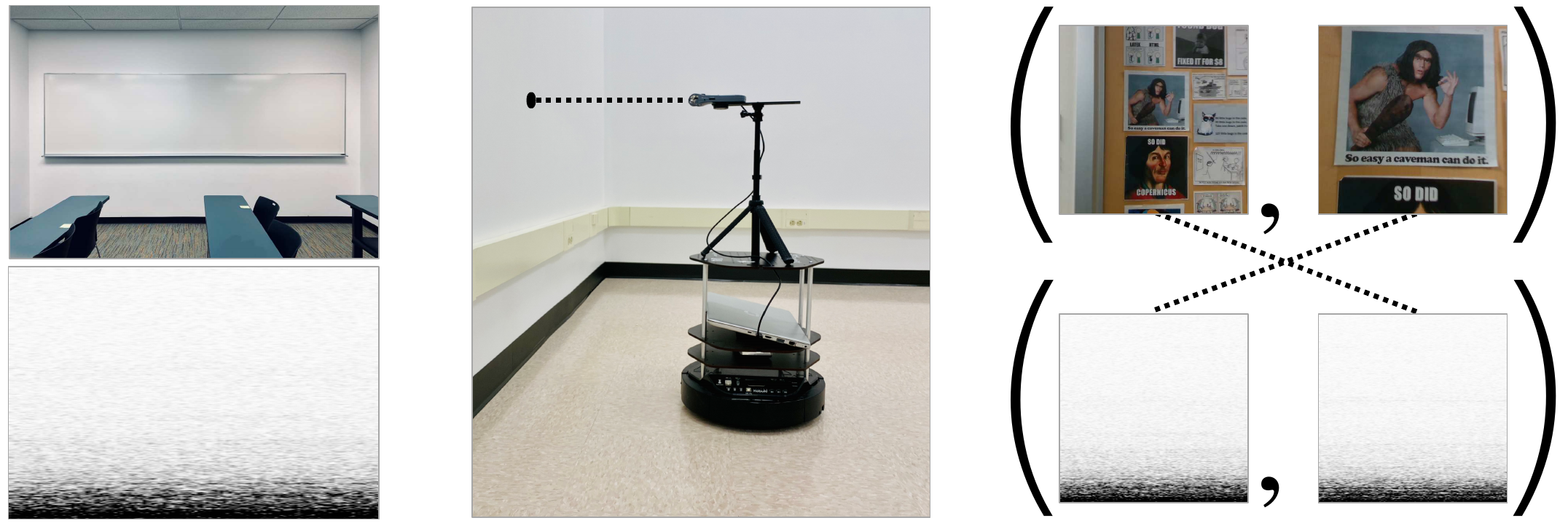}
\vspace{-6mm}
\begin{flushleft}
\hspace{-1.6mm} (a) {\em Quiet Campus} dataset \hspace{10.7mm} (b) Depth estimation  \hspace{13.5mm} (c) Multimodal self-supervision
\end{flushleft}

\captionof{figure}{What can ambient sound tell us about 3D scene structure? (a) We collect an ``in-the-wild'' dataset of paired audio and RGB-D recordings from quiet indoor scenes. (b) Given audio from a scene, we estimate distance to a wall. (c) We use this ambient sound to learn audio-visual representations through self-supervision.
}
\label{fig:teaser}
\end{center}%
{\let\thefootnote\relax\footnotetext{{\textsuperscript{*} Indicates equal contribution.}}}
\vspace{3mm} \begin{abstract}
From whirling ceiling fans to ticking clocks, the sounds that we hear subtly vary as we move through a scene.
We ask whether these ambient sounds convey information about 3D scene structure and, if so, whether they provide a useful learning signal for multimodal models. To study this, we collect a dataset of paired audio and RGB-D recordings from a variety of quiet indoor scenes. We then train models that estimate the distance to nearby walls, given only audio as input. %
We also use these recordings to learn multimodal representations through self-supervision, by training a network to associate images with their corresponding sounds. These results suggest that ambient sound conveys a surprising amount of information about scene structure, and that it is a useful signal for learning multimodal features.

\end{abstract}


\vspace{2mm} \section{Introduction}

Humans make extensive use of sound for understanding 3D scene structure. Recent work has sought to reproduce these capabilities in machines, often in simulations with loud, distinctive sounds, such as buzzing alarm clocks and footsteps. Yet it is unclear how often sounds like these are available to robots in practice. Real-world scenes tend to be quiet, and the sounds that we do hear within them, such as ventilation noise, tend to be ambiguous, and propagate through a scene in complex ways.

Human hearing, by contrast, is capable of estimating scene structure from subtle ambient sound cues. Perhaps the most familiar of these are contextual associations, such as knowing that noisy vents and sound-leaking windows tend to be attached to walls. Work in psychology, however, has also proposed that humans use simple low-level cues. These include using the accumulation of low frequencies to tell whether they are near walls~\citep{ashmead1998echolocation}, using sound shadows to estimate the shape of obstructions~\cite{rosenblum2007hearing}, and using reverberation to interpret scene geometry~\citep{traer2016statistics}. One feature of these cues is that they are available solely through {\em passive observation}. This is in contrast to echolocation, which requires the observers to actively produce sounds.

These sounds also provide a ``free" learning signal for vision~\cite{owens2016ambient}. Learning to successfully predict audio from images (or vice versa) requires understanding how the two modalities vary with pose and scene structure. For example, as one moves toward a sound-making object, it will simultaneously get louder and increase its size within the visual field. Rotating the camera, on the other hand, drastically alters the visual signal without significantly changing the sound. A model that can predict one modality from the other must learn both sources of variation, such as by learning a visual representation that is somewhat invariant to rotations but sensitive to distances to sound sources. %

In this paper, we ask whether ambient sounds that occur in {\em real-world scenes} convey 3D structure, and whether they can be used for multimodal self-supervised learning. %
To study this, we collected a dataset of ``in-the-wild'' audio recordings from quiet, indoor scenes typical of what a robot would encounter when solving navigation tasks. Each sound in our dataset is paired with a corresponding recording from an RGB-D sensor, which provides a visual signal and pseudo ground-truth depth. The resulting dataset, which we call the \emph{\audiodataset} dataset (Fig.~\ref{fig:teaser}a), covers a variety of room shapes, background sounds, and materials. 

Using this dataset, we conduct an experimental study of depth estimation from audio. First, we show that audio can be used to estimate the distance to nearby walls in a variety of scenarios, including predictions of the relative depth between two recordings. We demonstrate that the model can be used as part of a very simple robotic navigation system, in which a wheeled robot moves along a wall using ambient audio cues (Fig.~\ref{fig:teaser}b). %
Next, we ask whether ambient sound can be used as a learning signal, without explicit use of depth. We study several multimodal self-supervised learning formulations, each of which requires a model to associate subtle changes in sound with a visual signal (Fig.~\ref{fig:teaser}c). We show that the resulting model learns a feature set that can be used to solve downstream distance-to-wall estimation tasks.

We make the following contributions: i) a dataset of paired audio and RGB-D data from real-world scenes, ii) an experimental study of audio-based depth estimation, iii) showing that audio-visual recordings from these scenes can provide useful self-supervision for depth estimation tasks. Our results suggest that ambient sound conveys a surprising amount of information about structure, and that it provides a useful training signal for multimodal learning.

\section{Related Work}

\mypar{Human auditory perception.} The field of psychoacoustics has studied what humans can infer from sounds, such as size, material, shape, and depth. 
One well-known cue for depth estimation is {\em echolocation}, whereby humans estimate distance by generating sounds and analyzing their echoes~\citep{rosenzweig1955evidence, ashmead1989obstacle}. However, humans may also infer a great deal from passive observation. 
Our work is inspired by Ashmead \etal's studies of the navigation abilities of children with visual impairments~\citep{ashmead1998echolocation, ashmead1999auditory, ashmead2002lowfreqency}. They propose that visually impaired people use accumulations of low-frequency sounds to detect obstacles, and propose physical models of wave interference that would lead to changes in ambient sound fields near obstacles. This effect is similar to how a seashell held to one's ear amplifies ambient sounds to create ocean-like noises~\cite{chamberlain2006she}. Traer and McDermott~\citep{traer2016statistics} show that humans can use phase changes from natural reverberation to distinguish different spaces. While these cues have been studied extensively in psychoacoustics, they are not widely used in robotic perception and self-supervised learning.

\mypar{Audio-based depth estimation.}
We take inspiration from work in ocean acoustics that estimates the structure of the ocean from ambient sound using microphone arrays~\cite{sabra2005extracting,traer2011ocean}. In classic work, Kac~\cite{kac1966can} related the shape of a drum to its sound. Recently, Purushwalkam \etal~\cite{purushwalkam2020audio} reconstructed floor plans from audio-visual signals, using a simulated environment in which sound effects from FreeSound~\citep{font2013freesound} were synthetically inserted into the scenes (\eg, beeping alarm clocks, flushing toilets, and dishwashers). However, it is unclear how often distinctive sounds like these occur in practice, and whether the simplified models of sound propagation are sufficiently accurate. By contrast, we experimentally study audio-based depth estimation in real-world scenes. Other work uses multiple microphones to localize sound sources~\citep{thrun2005affine,gan2019self,yang2020telling,morgado2018self,chen2021boombox}. %
Our method, by contrast, requires only monaural audio.
Recent work also uses self-produced echolocation sounds produced by onboard speakers. Christensen~\etal~\citep{christensen2020batvision} predict depth maps from real-world scenes using echo responses. Gao~\etal~\citep{gao2020visualechoes} learns visual
representations by echolocation in a simulated environment~\citep{chen2020soundspaces}.
In contrast, we learn through passive observation, rather than active sensing. %

\mypar{Audio-visual learning.} 
Many recent works have proposed to use paired audio-visual data for representation learning. In seminal work, de Sa~\citep{de1994learning} proposed multimodal self-supervised learning as an alternative to single-modal models. Later, Ngiam~\etal~\citep{ngiam2011multimodal} learned a multimodal Boltzmann machine. Owens \etal~\citep{owens2016visually}  learned self-supervised visual representations from impact sounds, and used ambient sound to learn visual features~\citep{owens2016ambient}. In the latter, the sounds are taken from internet video and thus contain a much wider range of auditory events than what we consider in this work. %
Later work simultaneously learned audio and visual representations~\citep{arandjelovic2017look,owens2018learning,korbar2018cooperative,xiao2020audiovisual,morgado2021audio,asano2020labelling}.
Other work has learned cross-modal distillation~\citep{aytar2016soundnet}, sound source localization~\citep{senocak2018learning, arandjelovic2018objects, harwath2018jointly, owens2018learning, tian2018audio, afouras2020self, qian2020multiple, hu2020discriminative, chen2021localizing}, active speaker detection~\citep{chung2016out,chung2019perfect,roth2020ava}, source separation~\citep{gao2018learning, zhao2018sound, zhao2019sound, gao2021visualvoice}. 
We take inspiration from this work and show that quiet indoor sounds also provide a useful self-supervised learning signal.  %

\mypar{Visual depth prediction.} 
Recent work has learned to recover 3D structure from RGB images reconstructions~\citep{hoiem2007recovering,saxena2008make3d,eigen2015predicting, chen2016single, chen2019learning, li2018megadepth, godard2019digging}, via voxel grids, point clouds, and meshes~\citep{tulsiani2018factoring, gkioxari2019mesh, girdhar2016learning,yin2020learning}. 
Other work has made significant progress predicting relative camera pose~\citep{hartley2004mvg, hartley1997defense,wu20083d,qian2020associative3d,cai2021extreme}.
We take inspiration from the techniques proposed in these models, though we predict depth from audio instead of from images. %

\mypar{Robotic navigation.} %
Robots often use visual signals to navigate in novel environments~\citep{zhu2017target, mirowski2016learning, savva2019habitat, wijmans2019dd}. While vision is often a reliable cue for depth estimation, there are many situations where it is unavailable (e.g. due to sensor failures or poor lighting), necessitating ``backup" modalities.
Recent works have proposed methods that use sound for robot navigation. Those robotic systems are designed to localize sound sources and to navigate to audio goals in indoor environments~\citep{rascon2017localization, an2018reflection, chen2020soundspaces, gan2019look, chen2020learning, chen2020semantic}. Unlike these methods, which largely use distinctive sound sources, we use ambient sounds collected in real-world scenes.%

\begin{figure*}[!t]
    \vspace{-3.7em}
    \centering
    \hspace{-1.5em}
     \includegraphics[width=\textwidth]{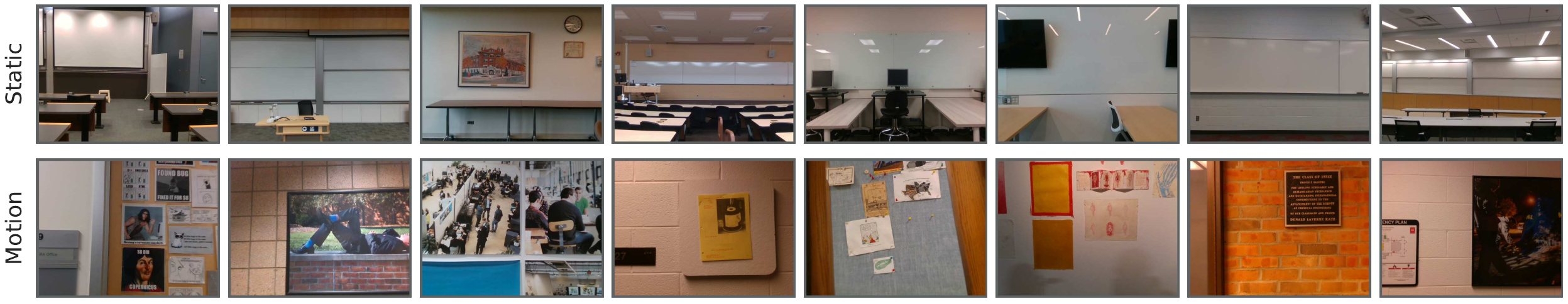}
    \caption{{\bf The {\em Quiet Campus} Dataset}. We collected a dataset of paired audio and RGB-D recordings from a variety of quiet indoor scenes. We show selected images from the \emph{\audioum} and \emph{\audiotd} subsets, which contain stationary and moving microphones respectively. Please refer to the project webpage for audio-visual examples. } 
    \label{fig:datasample}
\end{figure*}

\section{The {\em \audiodataset} Dataset}
\label{sec:data}

\newcommand{\numclassrooms}{46\xspace}
\newcommand{\numrecordings}{1300\xspace}
\newcommand{\numbuildings}{12\xspace}

\newcommand{\nummotionrooms}{46\xspace}
\newcommand{\nummotionrecordings}{1300\xspace}

\begin{wrapfigure}[16]{R}{0.15\textwidth}
\vspace{-5mm}
\hspace{-2.5mm}
\input{floats/fig_datacollect}
\end{wrapfigure}
To study the structure-from-audio problem, we collected a dataset containing a number of ``in-the-wild'' indoor ambient audio recordings, paired with concurrently recorded RGB-D data. We collected our data from a large number of classrooms and hallways of a college campus. To ensure that the sounds within the recordings were subtle ambient sounds (e.g. ventilation noise), rather than distinctive audio events (e.g. speech), we collected the data at times of the day when there were relatively few people occupying the buildings. %
Please refer to our webpage for a random sample of sounds.

We collected two types of data: {\em static} recordings in which the microphone and camera are stationary, and {\em motion} recordings where a human collector moves slowly toward or away from a wall. We show some visual examples from our collected dataset in Fig.~\ref{fig:datasample}. In our experiments, we split the data so that there is no physical overlap, such that the rooms in the training and tests sets are disjoint.

\mypar{{\capaudioum} recordings.} 
We collected data in \numclassrooms classrooms from \numbuildings buildings on The University of Michigan's campus, amounting to approximately 200 minutes audio. 
Inside each classroom, we selected $16-30$ positions and recorded 10 secs. of audio at each one. Since we are interested in predicting distance to walls, we point the camera and microphone toward the nearest wall when recording, so that the distance is well-defined. %
 
\mypar{{\capaudioum} recordings with dense coverage.} In order to evaluate generalization to different orientations, we use four classrooms from \emph{{\audioum}} recordings, divided each one into 20 to 35 grid cells of size $1.5 \times 1.5 \text{m}^2$ depending on their sizes, and collect 10 hours of audio densely with a wide range of angles (rather than always facing a wall). We call  this subset the \textit{{\audioum}-dense} recordings. 
These two datasets were recorded at different times of the year. The \textit{{\audioum}} recordings were made in spring 2021, when buildings were at reduced occupancy due to the COVID-19 pandemic, %
while  \textit{{\audioum}-dense} were recorded in the winter of 2020, when buildings were at normal occupancy. %

\mypar{{\capaudiotd} recordings.}
To help understand whether small motion (and the subtle changes in audio over space) provides useful information about structures, we collected approximately 90 minutes of videos in motion (222 videos total). To ensure that there are no unintended obstructions nearby, we collect this data in hallways.
During recording, the microphone and RGB-D camera move toward or away from a wall (Fig.~\ref{fig:datasample}).  As in the static setting, we point the microphone toward the wall. To reduce possible sources of sound, the human operator moves slowly and does not take footsteps. We split this dataset based on {\em building}, \ie, the videos in the training and test set are hallways recorded in different buildings.

\mypar{Estimating wall distance.} For experiments that require distance to the wall, we obtain this by cropping the center $320 \times 240$ region from the depth image, and averaging the depth values. Since the camera is oriented toward the wall, this is generally an accurate estimate of depth.

\mypar{Hardware.} We use a ZOOM H1n Handy Recorder for collecting audio since it is a simple, inexpensive, and widely-used stereo audio recorder. For the RGB-D data, we use Intel Real-Sense Depth Camera D415 (Fig.~\ref{fig:data_collect}).

\section{Predicting 3D Structure from Ambient Sound}
\label{sec:method}

We design models and experiments that will experimentally evaluate the ability of a model to learn to predict depth from audio in natural scenes.

\begin{figure*}[t]
    \vspace{-3.7em}
    \centering
    \includegraphics[width=\textwidth]{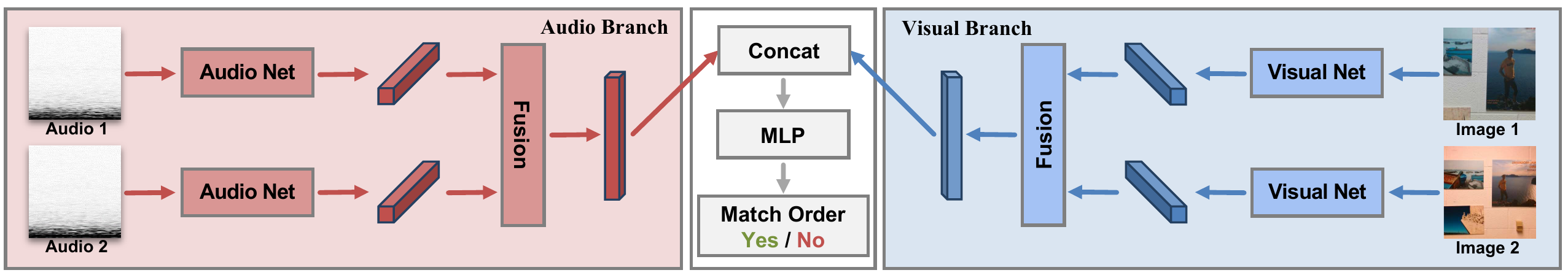}
    \caption{{\bf Self-supervised audio-visual model}. We encode a pair of visual and pair of audio examples using CNNs, and fuse each pair independently using concatenation. The two modalities are then combined and classified using a small multi-layer perceptron, which predicts whether the two pairs are in the same order. Please see \ref{appendix:self-supervision} for details.
    }
    \label{fig:method}
\end{figure*}

\subsection{Depth estimation tasks}
\label{sec:dep_est}
Inspired by work in human obstacle detection with sound~\citep{ashmead1989obstacle}, we study the ability of a machine hearing system to predict the distance to a wall ahead of them. We pose this problem in multiple ways, which has applications to a variety of robotic navigation applications.

\mypar{Obstacle detection.} First, we pose the problem of estimating whether the microphone is within a small distance of a wall (we use 0.5 meters), a binary classification problem. 

\mypar{Relative depth order.} In many cases, a robotic system needs only to move toward or away from an obstacle, and hence only needs to know which direction is closer to it. We also hypothesize that relative depth estimation is easier task than absolute estimation, as it is in visual depth perception~\citep{palmer1999vision,hoiem2007recovering}. Given two audio clips, we train a network to predict which one is closer to a wall. 

\mypar{Relative depth estimation.} Given two audio clips, we train a network to predict the difference between their distances to the wall. We also consider predicting the $\log$ of their difference, following the work in visual depth estimation~\citep{eigen2015predicting,saxena2008make3d}, and as a multi-way classification task, after clustering depth differences in the training set with $k$-means.%

\mypar{Absolute depth estimation.} Finally, we train a model to regress the distance to a wall, rather than a thresholded ``near-or-far'' prediction.

\subsection{Self-supervised audio-visual learning}

Having evaluated our model's ability to estimate depth directly from audio, we now ask whether we can learn about structure through audio-visual self-supervision. We hypothesize that, through the task of associating a raw~(RGB) visual signal with audio, we will learn a representation with features that will be helpful for downstream depth estimation tasks. 

One simple model, considered in prior work, is to solve a audio-visual synchronization task~\citep{chung2016out,owens2018learning,afouras2020self}: given the video, distinguish whether audio and visual stream have been misaligned through a random shift. We call this model the {\bf AV-Sync} model. 
Finally, we pose a new audio-visual learning problem that requires a model to learn {\em relative} depth between examples. We train a model to distinguish between {\em ordered} audio-visual pairs, which we call the {\bf AV-Order} model. Given a pair of audio tracks and a pair of images (which correspond to the two audio tracks), we train a network to determine whether the two pairs are in the same (or mismatched) order, which we frame as a classification problem.

\subsection{Audio-based robotic navigation}
To demonstrate the effectiveness of our depth estimation model, we use it to guide a wheeled robot (TurtleBot) to follow a wall. The robot uses the obstacle detection model to predict whether there is wall near the left/right of the robot, and moves accordingly. We use a simple policy, based on that of Gandhi \etal~\citep{gandhi2017learning} (please see \ref{appendix:robot} for full details). The robot moves forward in a sequence of steps. For each step, it rotates to an orientation that is derived from the obstacle detection model. It senses the audio on its left side, then rotates $180^\circ$ and senses audio to its right. We run the obstacle detection model on both audio recordings, and choose an orientation such that the robot moves in the direction that has lower probability of being near a wall. %

\begin{figure*}[!t]
    \vspace{-3.7em}
    \centering
    \includegraphics[width=\textwidth]{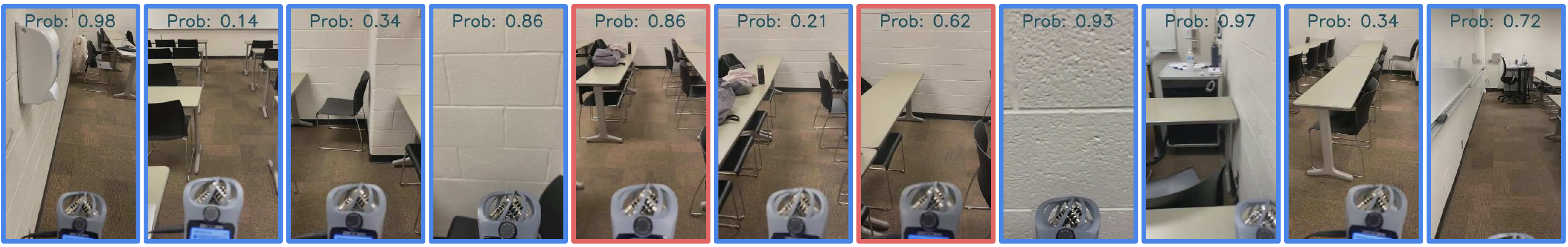}
    \caption{{\bf Obstacle detection}. After training our model, we walk through a room and show the predicted probability of being near a wall. The frames with \textcolor{blue}{blue} border are correct predictions while frames with \textcolor{red}{red} border are failure cases. Please refer the webpage for video results (with audio).} \label{fig:demo}
\end{figure*}

\subsection{Models}
\mypar{Network backbone.} 
We represent audio using the VGGish network~\citep{hershey2017cnn}. For our experiments, we either replace the final layer with a new regression or classification head, or we use the (128-dimensional) penultimate layer. The network takes 0.96 sec of audio as input, in the form of a log-mel spectrogram. Due to the importance of low-frequency sound in physical models of wall-sound interaction~\citep{ashmead2002lowfreqency}, we do not remove the very lowest frequencies of the audio, in contrast to common practice in speech and audio event recognition~\citep{hershey2017cnn}. We analyze the importance of this decision through experiments. We convert the stereo audio recordings to mono by averaging the two channels.

For models that use visual data, we use ResNet-18~\citep{he2016deep}, with  $224 \times 224$ images, and modify the last linear layer to output the feature vectors with the same dimension as audio features. We also consider a model whose weights are finetuned from ImageNet~\citep{deng2009imagenet}. %

\mypar{Learning details.} 
For fully supervised tasks, we use SGD~\citep{sutskever2013importance} optimizer with a learning rate $10^{-3}$, momentum $0.9$, and weight decay $5\times 10^{-4}$. For the self-supervised learning experiments, we use Adam~\citep{kingma2014adam} with a learning rate of $10^{-4}$ for representation learning. For linear probing experiments, we use SGD with a learning rate of $10^{-2}$. We schedule learning rates via cosine decay and pick the model with best validation performance. %

\mypar{Audio processing.} Our audio preprocessing resembles that of~\citep{hershey2017cnn}. We resample the audio to 16 kHz and extract segments of 0.96 secs. We then convert them to spectrograms by applying a Short-time Fourier transform at a stride of 10 ms and 25 ms, and convert them into $96 \times 64$ log-mel spectrograms. For augmentation, we randomly crop the waveform in time and multiply the amplitudes by a random factor uniformly sampled from [0.5, 1.5]. 

\mypar{Depth estimation networks.}
For the regression and binary classification tasks, we add a small multi-layer perceptron (MLP) on top of the VGGish network. For the relative estimation tasks, our network uses two audio inputs. For these tasks, we use a Siamese network that fuses the extracted feature vectors via concatenation prior to the MLP. For regression, we normalize depths to be in the range $[-0.5, 0.5]$ during training.

\mypar{Network for self-supervised learning tasks.} 
For AV-Sync task, we concatenate image and audio features before passing through the multi-layer perceptron. 
For the AV-Order model, we fuse Siamese networks from both modalities via concatenation and add a multi-layer perceptron (Fig.~\ref{fig:method}). In the downstream task, we take the output features from the last convolution layer after global average pooling and freeze them. We pass our 512-d learned features to a linear layer, which performs classification (and which is the only part of the model optimized during training).

\section{Experiments}
\label{sec:exp}

Having collected a dataset of in-the-wild audio and models for inferring structure from it, we conduct an experimental study of what can be inferred from ambient sound in real-world scenes. We report our results along with 95\% confidence intervals. 

\subsection{An experimental study of audio-based depth estimation}

\mypar{Obstacle detection.} First, we evaluate our model's ability to solve the obstacle detection problem (Tab.~\ref{tab:obstacle_with_rd}). We split the data such that the test rooms were not heard during training. We report accuracy on this binary classification task, as well as average precision (to account for possible dataset shift from using held-out rooms). We evaluated our model in both the static and motion settings. To help interpret the effectiveness of audio in the context of other modalities, we also included simple visual models, with both random initialization and ImageNet pretraining\footnote{We do not use a pretrained VGGish audio model since existing models remove low frequencies and are not compatible without significant changes.}. Interestingly, we see that the audio-based model performs better than chance in both cases, suggesting that ambient sound, indeed, conveys information about scene structure. 

We also evaluated our model's ability to generalize across orientations and recording times. We found that a model trained on \emph{\audioum} recordings could achieve 86.1\% AP and 72.9\% accuracy on \emph{\audioum-dense} recordings, indicating strong generalization performance. 

\begin{table}[tb]
\vspace{-5.0em}
    \centering
    \begin{minipage}[c][1\width]{0.35\textwidth}
    \captionsetup{type=figure}
        \centering
\includegraphics[width=\linewidth]{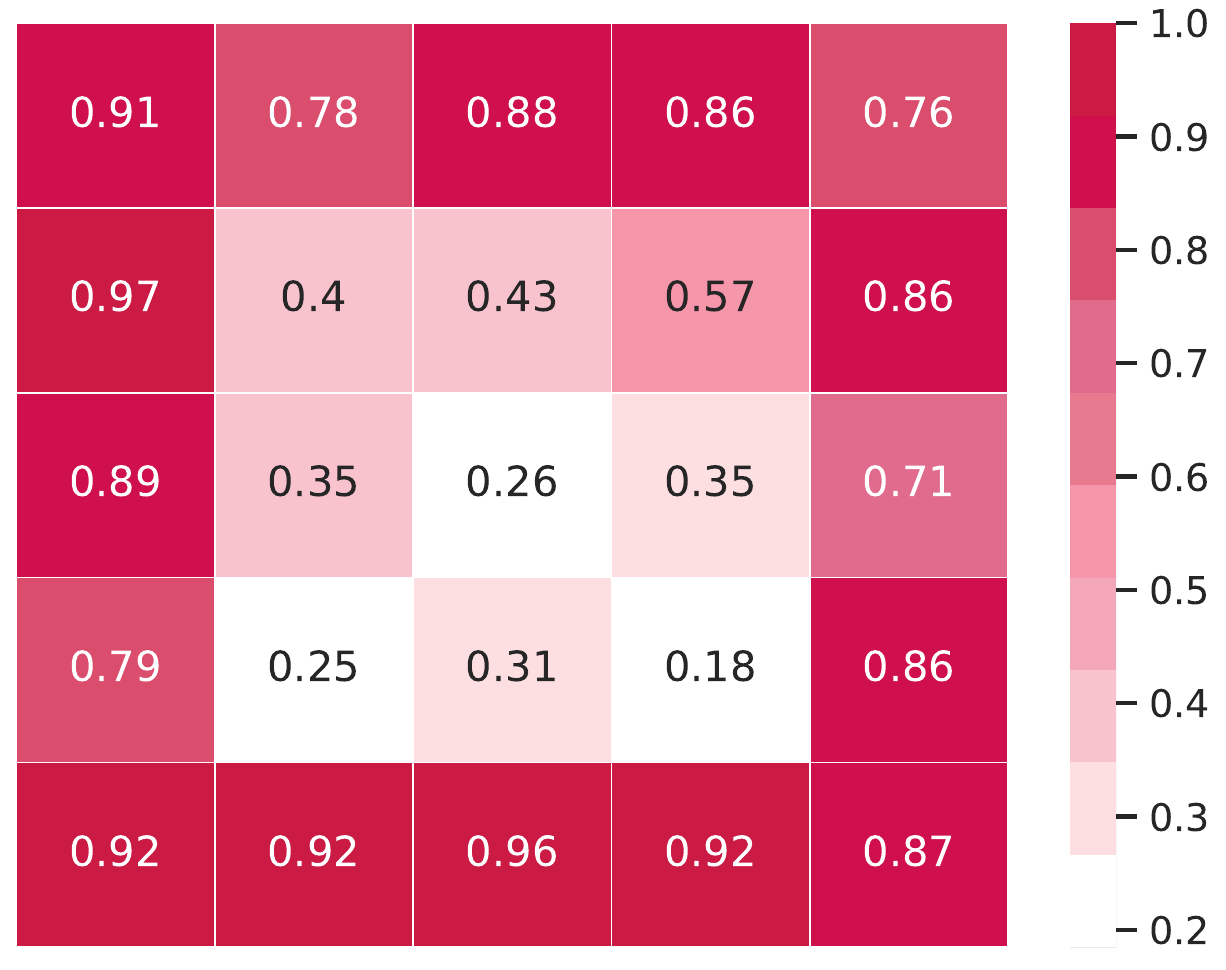}
\caption{{\bf Wall detection in a single room.}  We divide a room into a grid, and record audio within each cell. We show the average obstacle detection probability in each one. }
\label{fig:gridheatmap}

    \end{minipage}
    \hfill
    \begin{minipage}[c][1\width]{0.6\textwidth}
        \vspace{-1.7em}

\caption{{\bf Obstacle detection and relative depth order.} We evaluate our model's ability to determine whether a microphone is within 0.5 meters of a wall and identify which sound has a smaller distance to the wall. {\em Pre} refers to pretraining. %
}
  \label{tab:obstacle_with_rd}
  \centering
  \resizebox{1.0\columnwidth}{!}
  {
\begin{tabular}{lcccc|cc}
    \toprule
     &  &  & \multicolumn{2}{c}{Obstacle detection} & \multicolumn{2}{c}{Relative order} \\
    \midrule
    Model & Pre. & Task & AP(\%) & Acc(\%) & AP(\%) & Acc(\%) \\
    \midrule
    \ourrow Audio & &  {\audioum} &  68.3\ci{1.3} &   60.0\ci{0.9} &  85.5\ci{1.0} & 77.2\ci{0.8}\\
    Image &  & {\audioum}  & 99.2\ci{0.2} &  95.5\ci{0.5} &  94.6\ci{0.5} & 86.4\ci{0.7} \\
    Image & \checkmark & {\audioum}  & 99.5\ci{0.1} &  98.4\ci{0.4} & 97.7\ci{0.2} & 92.1\ci{0.5}\\
    Chance & & {\audioum}  & 46.4\ci{1.4} &  50.0\ci{1.0} & 47.2\ci{1.3} & 50.0\ci{1.0}\\
        
    \cmidrule(lr){1-7}%
    \ourrow Audio &  &  {\audiotd} &  65.6\ci{1.4} &   64.5\ci{0.9} & 87.1\ci{1.0} & 81.3\ci{0.8}\\
    Image  &  & {\audiotd} & 73.4\ci{1.2} &  68.2\ci{1.0}  & 87.9\ci{0.9} & 81.2\ci{0.8}\\
    Image  & \checkmark & {\audiotd} & 88.6\ci{0.7} &  78.5\ci{0.8} & 97.1\ci{0.3} & 90.6\ci{0.6}\\
    Chance & & {\audiotd} & 50.4\ci{1.4} &  50.0\ci{1.0} & 50.5\ci{1.4} & 50.0\ci{1.0} \\
    
    \bottomrule
  \end{tabular}
  }
    \end{minipage}
    \vspace{-4.0em}
\end{table}

\textls[0]{We qualitatively demonstrate the obstacle detection results by training a version of the model on the {\em \audioum-dense} data, since it contains a large number of viewpoint variations and angles (see \ref{appendix:static-dense} for quantitative results). We apply the learned obstacle detection model to every frame of a video in which a person walks through a room (Fig.~\ref{fig:demo}). In Fig.~\ref{fig:gridheatmap}, we also apply this model densely within a room, divided according to its spatial grid. We averaged the probability of the model for each grid cell. We see that it successfully distinguishes between the boundaries of the room and the center.}

\begin{wraptable}{r}{0.65\textwidth}
\vspace{-1.4em}

  \caption{{\bf Relative depth ratio.} We evaluate our model's ability of predicting relative depth ratio from two ambient sounds, for the \emph{\audiotd} recordings.}
  \label{tab:rdratio}
  \centering
 \resizebox{1.0\linewidth}{!}
 {
  \begin{tabular}{lccccccc}
    \toprule
     & \multicolumn{3}{c}{Regression} & \multicolumn{3}{c}{Regression-by-Classification} \\
    \midrule
    Model  & MAE $\downarrow$    & Med. $\downarrow$  & $R^{2}$ $\uparrow$  & Top-1 $\uparrow$  & Top-5 $\uparrow$  & Avg. Dist $\downarrow$ \\ 
    \midrule
    \ourrow Audio     &  0.55\ci{.01}  &  0.44\ci{.01}  &  0.48\ci{.02} &  22.8\ci{0.8}  &  80.7\ci{0.7}  & 1.66\ci{.03}\\
    Image  &   0.54\ci{.01}  & 0.42\ci{.01}  & 0.49\ci{.02} & 26.6\ci{0.8}  & 83.6\ci{0.7}  & 1.47\ci{.02}\\
    Image~(Pre.)  & 0.39\ci{.01}  & 0.29\ci{.01}  & 0.72\ci{.01} & 34.2\ci{0.9}  & 90.5\ci{0.6}  & 1.15\ci{.02}\\
    Chance &    0.89\ci{.01}  & 0.79\ci{.01}  & 0.00\ci{.00} & 9.45\ci{0.6}  & 52.6\ci{1.0}  & 2.79\ci{.04}\\
    No input     & 0.82\ci{.01}  & 0.75\ci{.01}  & 0.00\ci{.00} & 10.7\ci{0.6}  & 51.6\ci{1.0}  & 4.50\ci{.05}\\
    \bottomrule
  \end{tabular}
 }

\vspace{-1.5em}
\end{wraptable} 
\mypar{Relative depth order.} In Tab.~\ref{tab:obstacle_with_rd}, we evaluate our model's ability to predict which of two audio clips is closer to a wall. We find that the model obtains results that are significantly better than chance, and that relative prediction is significantly easier than obstacle detection. We also asked whether our model might use binaural cues (even though the stereo channels are averaged). Using only the left channel, we obtain nearly identical performance: 87.2\% AP and 81.2\% accuracy. Finally, we analyzed the effects of mixing in spurious distractor sounds (see \ref{appendix:nonambient}). %

\begin{wraptable}{r}{0.65\textwidth}
\vspace{-1.3em}

\centering
 \caption{{\bf Absolute depth estimation.} We evaluate our model's ability of  predicting absolute distance to the wall for the \emph{\audiotd} recordings. }
  \label{tab:abs}
 \resizebox{1\linewidth}{!}
 { 
  \begin{tabular}{clccccccc}
    \toprule
        &  & \multicolumn{3}{c}{Regression} & \multicolumn{3}{c}{Regression-by-Classification} \\
    \midrule
    & Model  & MAE $\downarrow$    & Med. $\downarrow$  & $R^{2}$ $\uparrow$ & Top-1 $\uparrow$  & Top-5 $\uparrow$  & Avg. Dist $\downarrow$ \\ 
    \midrule
    \parbox[c]{2mm}{\multirow{4}{*}{\rotatebox[origin=c]{90}{Single}}} 
     &  \ourcell Audio  & \ourcell 0.28\ci{.00}  & \ourcell  0.25\ci{.01}  &  \ourcell -0.34\ci{.03} & \ourcell 30.8\ci{0.9}  & \ourcell 88.3\ci{0.6}  & \ourcell 1.11\ci{.02}\\
    & Image &   0.31\ci{.00}  & 0.27\ci{.01}  & -0.67\ci{.07} & 35.6\ci{0.9}  & 95.9\ci{0.4}  & 1.05\ci{.02}\\
    & Image~(Pre.) &  0.26\ci{.00}  & 0.21\ci{.01}  & -0.24\ci{.04} & 50.8\ci{1.0}  & 99.2\ci{0.2}  & 0.62\ci{.01}\\
    & No input   & 0.28\ci{.00}  & 0.27\ci{.01}  & -0.19\ci{.02} & 24.3\ci{0.8}  & 88.3\ci{0.6}  & 1.07\ci{.01}\\
    \cmidrule(lr){2-9}
     \parbox[c]{2mm}{\multirow{4}{*}{\rotatebox[origin=c]{90}{Conditional}}} 
     & \ourcell Audio   & \ourcell 0.21\ci{.00}  & \ourcell 0.17\ci{.00}  & \ourcell 0.19\ci{.02} & \ourcell 36.9\ci{1.0}  &  \ourcell 90.0\ci{.06}  & \ourcell 1.17\ci{.02} \\
    & Image   & 0.22\ci{.00}  & 0.18\ci{.00}  & 0.12\ci{.02}  & 38.2\ci{0.9}  & 95.5\ci{0.4}  &  0.93\ci{.02}\\
    & Image~(Pre.)   & 0.18\ci{.00}  & 0.14\ci{.00}  & 0.39\ci{.02} & 51.7\ci{1.0}  & 99.8\ci{0.1}  & 0.59\ci{.01}\\
    & No input     & 0.25\ci{.00}  & 0.23\ci{.00}  & 0.01\ci{.01} & 26.4\ci{0.8}  & 95.9\ci{0.4}  & 1.43\ci{.03}\\
    \cmidrule(lr){2-9}
    & Chance   & 0.78\ci{.01}  & 0.84\ci{.01}  & -3.38\ci{0.23} &  23.3\ci{1.2} & 56.9\ci{0.9}  & 2.83\ci{.02} \\
    \bottomrule
  \end{tabular}
  }

\vspace{-1.3em}
\end{wraptable} 
\mypar{Relative depth ratio.} Since the network is able to tell relative depth order well, we explore relative depth setup further by predicting depth difference of audio pairs~(log depth ratio). We use regression to predict the log-ratio from ambient sounds with an L1 loss and via classification. We report the performance of mean absolute error, median error, and $R^2$. For classification, we report top-1 and top-5 accuracy, and report average distance from the ground truth bin using the distance between centers of paired bins. We report the performance of zeroing out two inputs as a baseline. As shown in Tab.~\ref{tab:rdratio}, we can see that our method outperforms chance and, interestingly, has a relatively small gap between the vision-based approach.%

\mypar{Absolute depth estimation.} We evaluated our model's ability to predict depth from a single audio example, using similar experiments as the relative depth ratio prediction (Tab.~\ref{tab:abs}). %
Our model is more accurate than chance (which we obtain for all metrics by simply ablating the input and training). Likewise, the regression-by-classification model improves over chance performance. 

We also consider another model %
that estimates depth for a given audio input after conditioning on both a {\em reference sound} recorded within the same scene and its depth (see \ref{appendix:conditional_ade} for details). 
We call this the {\em conditional} absolute depth model. This model obtains better performance than our unconditional model, suggesting that the reference audio provides  useful information.

\begin{table}[tb]
\vspace{-3.5em}
    \centering
    \begin{minipage}[c][1\width]{0.33\textwidth}
    \captionsetup{type=figure}
        \centering
\includegraphics[width=1\linewidth]{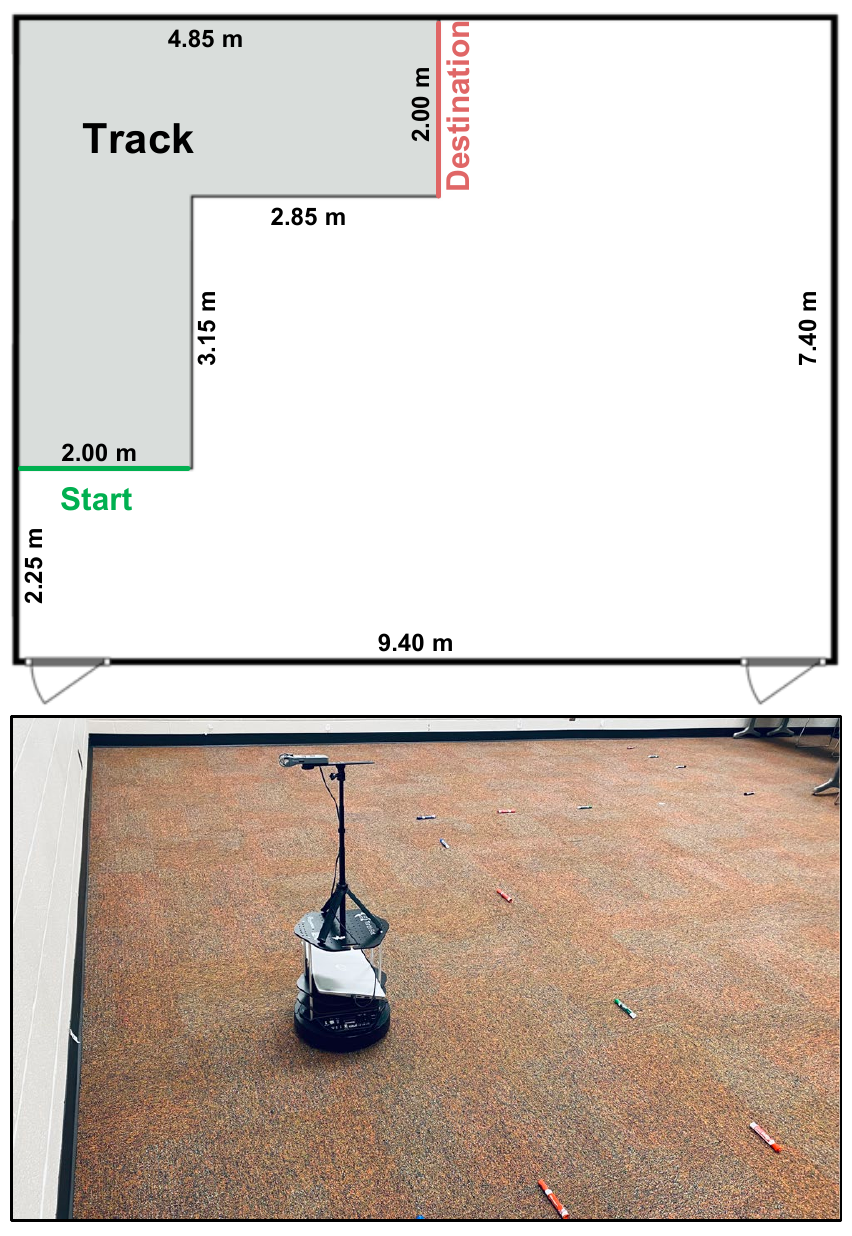}
\caption{{Classroom floor plan and track setting.} }
\label{fig:floorplan}

    \end{minipage}
    \hfill
    \begin{minipage}[c][1\width]{0.6\textwidth}
        \vspace{-1.5em}
         \centering
 \caption{{\bf Linear probing experiments.} We evaluate our self-supervised feature set for {\bf obstacle detection} and {\bf relative depth order}, for the \emph{\audiotd} recordings. Here, {\em Audio} means taking audio only as inputs. {\em Visual} means taking images only as inputs. {\em Both} means taking both audio and image as inputs.}
 \label{tab:avco_binary}
\resizebox{1.02\linewidth}{!}
{
  \begin{tabular}{clccc@{\hskip20pt}cc} 
    \toprule
       & &  & \multicolumn{2}{c@{\hskip20pt}}{Obstacle detection} & \multicolumn{2}{c}{Relative order} \\
    \midrule
    & Model & Pre. & AP(\%) & Acc(\%) & AP(\%) & Acc(\%) \\
    \midrule
    \parbox[c]{2mm}{\multirow{8}{*}{\rotatebox[origin=c]{90}{Audio}}}
    & Scratch &  & 61.9\ci{1.5} &  60.3\ci{0.9} & 78.0\ci{1.4} &  73.1\ci{0.9}\\
    & VGGish~\cite{hershey2017cnn} &  & 58.2\ci{1.3} &  56.0\ci{1.0} & 61.1\ci{1.4} &  61.2\ci{1.0}\\
    & AV-Sync &  & \textbf{69.1}\ci{1.4} &  \textbf{64.0}\ci{0.9} & 80.2\ci{1.3} &  74.1\ci{0.8} \\
    & AV-Order &  & 63.4\ci{1.4} &  61.5\ci{0.9} & \textbf{84.2}\ci{1.2} &  \textbf{79.4}\ci{0.7} \\
    \cdashlinelr{2-7}
    & VGGish~\cite{hershey2017cnn} & \checkmark & 59.0\ci{1.5} &  56.7\ci{1.0} & 67.7\ci{1.4} &  64.5\ci{0.9}  \\
    & AV-Sync & \checkmark & \textbf{65.3}\ci{1.4} &  \textbf{62.8}\ci{0.9} & 82.1\ci{1.2} &  76.4\ci{0.8} \\
    & AV-Order & \checkmark & 62.8\ci{1.5} &  64.5\ci{0.9} & \textbf{85.5}\ci{1.1} &  \textbf{80.7}\ci{0.8} \\
    \midrule
    \parbox[c]{2mm}{\multirow{6}{*}{\rotatebox[origin=c]{90}{Visual}}}
    & Scratch & & 70.1\ci{1.3} &  64.0\ci{0.9} & 79.7\ci{1.1} &  71.5\ci{0.9} \\
    & AV-Sync &   & \textbf{77.1}\ci{1.1} &  \textbf{69.2}\ci{0.9} & 85.3\ci{0.9} &  76.1\ci{0.8} \\
    & AV-Order &  & 76.8\ci{1.1} &  68.8\ci{0.9} & \textbf{87.4}\ci{0.9} & \textbf{79.1}\ci{0.8} \\
    \cdashlinelr{2-7}
    & ImageNet~\cite{deng2009imagenet, he2016deep}& \checkmark & 80.4\ci{1.2} &  74.5\ci{0.8} & 94.0\ci{0.5} &  85.8\ci{0.7}  \\
    & AV-Sync & \checkmark & \textbf{89.0}\ci{0.8} &  75.6\ci{0.8} & 92.8\ci{0.6} &  85.4\ci{0.7} \\
    & AV-Order & \checkmark & 86.5\ci{1.1} &  \textbf{76.3}\ci{0.8} & \textbf{95.8}\ci{0.4} &  \textbf{88.9}\ci{0.6} \\
    \midrule
    \parbox[c]{2mm}{\multirow{2}{*}{\rotatebox[origin=c]{90}{Both }}}
    & AV-Order & & 77.1\ci{1.1}  & 69.1\ci{0.9}   & 89.0\ci{0.8}  & 80.8\ci{0.8}   \\
     & AV-Order &\checkmark &  88.1\ci{0.9} & 76.9\ci{0.8}   &  95.8\ci{0.4} & 88.9\ci{0.6}   \\
    \bottomrule
  \end{tabular}
  }

    \end{minipage}
    \vspace{-1.7em}
\end{table}

\subsection{Audio-visual representation learning}
\begin{wraptable}{r}{0.6\textwidth}
\vspace{-1.7em}
\hspace{-2.5mm}
 \centering
\caption{{\bf Linear probing experiments.} We evaluate our learned representation for {\bf relative depth ratio} for the \emph{\audiotd} recordings.}
  \label{tab:avco_multi}
\resizebox{1\linewidth}{!}
 {
  \begin{tabular}{clcccc}
    \toprule
    & Model & Pre.   & Top-1 (\%) $\uparrow$  & Top-5 (\%) $\uparrow$  & Avg. Dist $\downarrow$ \\ 
    \midrule
    \parbox[c]{2mm}{\multirow{7}{*}{\rotatebox[origin=c]{90}{Audio}}}
    & Scratch &  & 19.2\ci{0.8}  & 72.8\ci{0.8}  & 2.33\ci{0.04} \\
    & VGGish~\cite{hershey2017cnn}&   & 14.4\ci{0.7}  & 53.9\ci{1.0}  & 3.78\ci{0.06} \\
    & AV-Sync. &  & 19.2\ci{0.7}  & 72.7\ci{0.8}  & 2.07\ci{0.03} \\
    & AV-Order &  & \textbf{22.2}\ci{0.9}  & \textbf{79.6}\ci{0.8}  & \textbf{1.86}\ci{0.03} \\
    \cdashlinelr{2-6}
    & VGGish~\cite{hershey2017cnn} & \checkmark & 15.6\ci{0.7}  & 54.0\ci{1.0}  & 3.59\ci{0.05} \\
    & AV-Sync. & \checkmark & 20.7\ci{0.8}  & 75.1\ci{0.9}  & 1.99\ci{0.03} \\
    & AV-Order & \checkmark  & \textbf{23.6}\ci{0.9}  & \textbf{80.5}\ci{0.7}  & \textbf{1.75}\ci{0.03} \\
    \midrule
    \parbox[c]{2mm}{\multirow{6}{*}{\rotatebox[origin=c]{90}{Visual}}}
    & Scratch  & & 18.5\ci{0.8}  & 70.8\ci{0.8}  & 2.66\ci{0.05} \\
    & AV-Sync. &  & 22.2\ci{0.8}  &  76.8\ci{0.8}  & 1.85\ci{0.03} \\
    & AV-Order  &  & \textbf{24.7}\ci{0.8}  & \textbf{80.2}\ci{0.8}  & \textbf{1.71}\ci{0.03} \\
    \cdashlinelr{2-6}
    & ImageNet~\cite{deng2009imagenet, he2016deep} & \checkmark  & 27.4\ci{0.9}  & 87.1\ci{0.7} & 1.60\ci{0.03} \\
    & AV-Sync. & \checkmark & 27.5\ci{0.8}  & 85.2\ci{0.7}  & 1.53\ci{0.03} \\
    & AV-Order & \checkmark   & \textbf{28.9}\ci{0.9}  & \textbf{88.6}\ci{0.6}  & \textbf{1.40}\ci{0.03} \\
    \midrule
    \parbox[c]{2mm}{\multirow{2}{*}{\rotatebox[origin=c]{90}{Both }}}
    & AV-Order  &   & 23.8\ci{0.8}  &  81.5\ci{0.7}    &   1.59\ci{0.03}      \\
    & AV-Order  & \checkmark  & 30.0\ci{0.9}  & 89.3\ci{0.6}     &  1.31\ci{0.03}    \\
    \bottomrule
  \end{tabular}
}

\vspace{-1.3em}
\end{wraptable} 
Finally, we investigate whether we learn useful representation with \textbf{AV-Order} and \textbf{AV-Sync} models. In the AV-Order binary classification pretext task, our model obtains 85.8\% AP and 75.8\% accuracy (80.3\% AP and 70.7\% accuracy with random initialization) on the test set where the random chance is 50\%. 
Given its performance on the pretext task, we test whether the learned audio and visual representations are useful for the downstream depth-estimation tasks: (1) obstacle detection and (2) relative depth estimation. 
To evaluate the quality of our spatial representation, we freeze the learned features and train a linear classifier on the depth estimation tasks (Section~\ref{sec:dep_est}). 
In the downstream tasks, we evaluate image and audio features individually and compare them with random features, pre-trained ImageNet features~\citep{deng2009imagenet, he2016deep}, pre-trained AudioSet features~\citep{hershey2017cnn} and \textbf{AV-Sync} features. Since AudioSet pre-trained weights require a special audio pre-processing that differs from ours, we provide another random feature baseline based on the same processing as ~\citep{hershey2017cnn}.

\mypar{Relative depth.} We report results for relative depth order and relative depth ratio in Tab.~\ref{tab:avco_binary} and Tab.~\ref{tab:avco_multi} respectively. Our method outperforms baselines in both audio and visual modalities suggesting that we learn useful depth representations. Moreover, our results are often close to supervised features. 

\mypar{Obstacle detection.} As the results shown in Tab.~\ref{tab:avco_binary}, AV-Order model outperforms Scratch and AudioSet pretraining, while it loses to AV-Sync model. The  AV-Order obtains worse performance since AV-Order, perhaps because the model deals with two inputs during the pretext task. We provide an additional experiment by fine-tuning our learned AV-Order feature and we compare with the fully supervised method in Tab.~\ref{tab:obstacle_with_rd}. This obtains a 4\% AP boost (70.0\% v.s. 65.6\%), suggesting that our learned representation is helpful. 

Interestingly, the multimodal models that use {\em both} audio and visual features consistently outperform the single-modality models, suggesting the audio is useful in conjunction with a visual signal. We conduct few-shot learning experiments to show audio-visual self-supervision improves fine-tuning performance (please see \ref{appendix:self-supervision} for the details).

\label{sec:audio_analysis}
\begin{wrapfigure}[14]{R}{0.45\textwidth}
\vspace{-5mm}
\hspace{-2.5mm}
    \input{floats/fig_freq.tex}
\end{wrapfigure}
\mypar{Audio cues.}
To help understand what information the model is using, we retrained our model with access to only part of the frequency spectrum. We zeroed out different frequency ranges in the Mel spectrogram, then re-trained the relative depth order model.
As can be seen in Fig.~\ref{fig:freq}, the accuracy of the model with low frequencies~(0-1000Hz) is comparable to the model with the full frequency range, suggesting that our model mainly use those frequencies to infer the structure information.
This may be because there is relatively little high frequency information in the dataset. 
We also compare the data distribution of our dataset with (non-ambient) music sounds from the VGGSound dataset~\cite{chen2020vggsound} (see \ref{appendix:vggsound} for details).

\subsection{Robotic navigation with ambient sound}

\begin{wrapfigure}[11]{R}{0.4\textwidth}
\vspace{-6mm}
\hspace{-2.5mm}
    \input{floats/fig_robot_exp.tex}
\end{wrapfigure}
We ask whether our depth estimation model can be used to guide a robot to follow a wall, solely using ambient sound. For this, we used a model trained on \emph{{\audioum}} recordings, and evaluate on a classroom that was not included in the training set. The track is 8m long, and forces the robot to turn a corner (Fig.~\ref{fig:floorplan}). The robot starts navigating from one of several unknown locations (40cm, 80cm, or 120cm from the wall) and orientations ($30^{\circ}$ left or right, or facing forward). We repeat the experiment 18 times and measure the average distance along the track it attained before colliding with a boundary. Quantitative results are shown in Fig.~\ref{fig:navig}. We compare with a random policy, a ``straight line'' policy that always drives forward, and a policy based on the vision model. For the random policy, the setting is the same as our method, except that we use an obstacle detection network with random weights. The straight line policy is effective when the robot starts with a front-facing orientation. It also has the advantage that it does not rotate to sense audio, and hence avoids accumulating errors\footnote{These errors could be reduced by using a multi-microphone array, though for simplicity we use a single microphone.}. %
As expected, the visual model significantly outperforms the audio model. We see that our method performs better than random and straight-line policies, and successfully guides the robot past the corner in nearly half of the trials. We repeat experiments in the different rooms (please see \ref{appendix:robot} for additional results).

\section{Conclusion}

We see this work as opening two directions. The first is in creating robotic systems that exploit real-world, passively obtained audio signals for depth estimation, especially as an inexpensive ``backup" signal that can be used when other modalities fail. %
Second, we see our work as a step toward using inexpensive and plentiful audio signals to help train other modalities, such as vision.%

\mypar{Limitations.} While ambient sound provides useful information about scene structure, it is generally less effective than vision. We also lack a full understanding of what cues the model is using, and what can (and cannot) be perceived from it, such as whether there are fundamental limitations on resolution, material, or angle. Additionally, our experiments were conducted on a single college campus, and hence may not be representative of all indoor spaces.

\clearpage

\acknowledgments{ We thank James Traer for his invaluable help explaining work on human perception of ambient sound. 
We would also like to thank Linyi Jin and Shengyi Qian for help with visual experiment setups and their valuable suggestions. We also thank David Fouhey for his comments and feedback, Peter Gaskell for his help on robot equipment, and Yichen Yang for the help on data collection. We thank the valuable suggestions and feedbacks from David Harwath, Kristen Grauman and UT-Austin Computer Vision Group. This work was funded in part by DARPA Semafor and Cisco Systems. The views, opinions and/or findings expressed are those of the authors and should not be interpreted as representing the official views or policies of the Department of Defense or the U.S. Government.} %

\bibliography{egbib}  %

\begin{thebibliography}{82}
\providecommand{\natexlab}[1]{#1}
\providecommand{\url}[1]{\texttt{#1}}
\expandafter\ifx\csname urlstyle\endcsname\relax
  \providecommand{\doi}[1]{doi: #1}\else
  \providecommand{\doi}{doi: \begingroup \urlstyle{rm}\Url}\fi

\bibitem[Ashmead et~al.(1998)Ashmead, Wall, Eaton, Ebinger, Snook-Hill, Guth,
  and Yang]{ashmead1998echolocation}
D.~H. Ashmead, R.~S. Wall, S.~B. Eaton, K.~A. Ebinger, M.-M. Snook-Hill, D.~A.
  Guth, and X.~Yang.
\newblock Echolocation reconsidered: Using spatial variations in the ambient
  sound field to guide locomotion.
\newblock \emph{Journal of Visual Impairment \& Blindness}, 92\penalty0
  (9):\penalty0 615--632, 1998.

\bibitem[Rosenblum and Robart(2007)]{rosenblum2007hearing}
L.~D. Rosenblum and R.~L. Robart.
\newblock Hearing silent shapes: Identifying the shape of a sound-obstructing
  surface.
\newblock \emph{Ecological Psychology}, 19\penalty0 (4):\penalty0 351--366,
  2007.

\bibitem[Traer and McDermott(2016)]{traer2016statistics}
J.~Traer and J.~H. McDermott.
\newblock Statistics of natural reverberation enable perceptual separation of
  sound and space.
\newblock \emph{Proceedings of the National Academy of Sciences}, 113\penalty0
  (48):\penalty0 E7856--E7865, 2016.

\bibitem[Owens et~al.(2016)Owens, Wu, McDermott, Freeman, and
  Torralba]{owens2016ambient}
A.~Owens, J.~Wu, J.~H. McDermott, W.~T. Freeman, and A.~Torralba.
\newblock Ambient sound provides supervision for visual learning.
\newblock In \emph{ECCV}. Springer, 2016.

\bibitem[Rosenzweig et~al.(1955)Rosenzweig, Riley, and
  Krech]{rosenzweig1955evidence}
M.~R. Rosenzweig, D.~A. Riley, and D.~Krech.
\newblock Evidence for echolocation in the rat.
\newblock \emph{Science}, 1955.

\bibitem[Ashmead et~al.(1989)Ashmead, Hill, and Talor]{ashmead1989obstacle}
D.~H. Ashmead, E.~W. Hill, and C.~R. Talor.
\newblock Obstacle perception by ongenitally blind children.
\newblock \emph{Perception \& psychophysics}, 46\penalty0 (5):\penalty0
  425--433, 1989.

\bibitem[Ashmead and Wall(1999)]{ashmead1999auditory}
D.~H. Ashmead and R.~S. Wall.
\newblock Auditory perception of walls via spectral variations in the ambient
  sound field.
\newblock \emph{Journal of rehabilitation research and development},
  36\penalty0 (4):\penalty0 313—322, October 1999.
\newblock ISSN 0748-7711.

\bibitem[Ashmead and Wall(2002)]{ashmead2002lowfreqency}
D.~H. Ashmead and R.~S. Wall.
\newblock Low frequency sound as a navigational tool for people with visual
  impairments.
\newblock \emph{Journal of Low Frequency Noise, Vibration and Active Control},
  21\penalty0 (4):\penalty0 199--205, 2002.
\newblock \doi{10.1260/026309202321834645}.

\bibitem[Chamberlain et~al.(2006)Chamberlain, Rosenblum, and
  Robart]{chamberlain2006she}
E.~J. Chamberlain, L.~D. Rosenblum, and R.~L. Robart.
\newblock She hears seashells: Detection of small resonant cavities via ambient
  sound.
\newblock \emph{The Journal of the Acoustical Society of America}, 2006.

\bibitem[Sabra et~al.(2005)Sabra, Gerstoft, Roux, Kuperman, and
  Fehler]{sabra2005extracting}
K.~G. Sabra, P.~Gerstoft, P.~Roux, W.~Kuperman, and M.~C. Fehler.
\newblock Extracting time-domain green's function estimates from ambient
  seismic noise.
\newblock \emph{Geophysical Research Letters}, 32\penalty0 (3), 2005.

\bibitem[Traer et~al.(2011)Traer, Gerstoft, and Hodgkiss]{traer2011ocean}
J.~Traer, P.~Gerstoft, and W.~S. Hodgkiss.
\newblock Ocean bottom profiling with ambient noise: A model for the passive
  fathometer.
\newblock \emph{The Journal of the Acoustical Society of America}, 129\penalty0
  (4):\penalty0 1825--1836, 2011.

\bibitem[Kac(1966)]{kac1966can}
M.~Kac.
\newblock Can one hear the shape of a drum?
\newblock \emph{The american mathematical monthly}, 73\penalty0 (4P2):\penalty0
  1--23, 1966.

\bibitem[Purushwalkam et~al.(2020)Purushwalkam, Gari, Ithapu, Schissler,
  Robinson, Gupta, and Grauman]{purushwalkam2020audio}
S.~Purushwalkam, S.~V.~A. Gari, V.~K. Ithapu, C.~Schissler, P.~Robinson,
  A.~Gupta, and K.~Grauman.
\newblock Audio-visual floorplan reconstruction.
\newblock \emph{arXiv preprint arXiv:2012.15470}, 2020.

\bibitem[Font et~al.(2013)Font, Roma, and Serra]{font2013freesound}
F.~Font, G.~Roma, and X.~Serra.
\newblock Freesound technical demo.
\newblock In \emph{ACM International Conference on Multimedia
  (MM{\textquoteright}13)}, pages 411--412, Barcelona, Spain, 21/10/2013 2013.
  ACM, ACM.
\newblock ISBN 978-1-4503-2404-5.
\newblock \doi{10.1145/2502081.2502245}.

\bibitem[Thrun(2005)]{thrun2005affine}
S.~Thrun.
\newblock Affine structure from sound.
\newblock \emph{Advances in Neural Information Processing Systems},
  18:\penalty0 1353--1360, 2005.

\bibitem[Gan et~al.(2019)Gan, Zhao, Chen, Cox, and Torralba]{gan2019self}
C.~Gan, H.~Zhao, P.~Chen, D.~Cox, and A.~Torralba.
\newblock Self-supervised moving vehicle tracking with stereo sound.
\newblock In \emph{Proceedings of the IEEE/CVF International Conference on
  Computer Vision}, pages 7053--7062, 2019.

\bibitem[Yang et~al.(2020)Yang, Russell, and Salamon]{yang2020telling}
K.~Yang, B.~Russell, and J.~Salamon.
\newblock Telling left from right: Learning spatial correspondence of sight and
  sound.
\newblock In \emph{Proceedings of the IEEE/CVF Conference on Computer Vision
  and Pattern Recognition}, pages 9932--9941, 2020.

\bibitem[Morgado et~al.(2018)Morgado, Vasconcelos, Langlois, and
  Wang]{morgado2018self}
P.~Morgado, N.~Vasconcelos, T.~Langlois, and O.~Wang.
\newblock Self-supervised generation of spatial audio for 360 video.
\newblock \emph{arXiv preprint arXiv:1809.02587}, 2018.

\bibitem[Chen et~al.(2021)Chen, Chiquier, Lipson, and
  Vondrick]{chen2021boombox}
B.~Chen, M.~Chiquier, H.~Lipson, and C.~Vondrick.
\newblock The boombox: Visual reconstruction from acoustic vibrations.
\newblock \emph{arXiv preprint arXiv:2105.08052}, 2021.

\bibitem[Christensen et~al.(2020)Christensen, Hornauer, and
  Stella]{christensen2020batvision}
J.~H. Christensen, S.~Hornauer, and X.~Y. Stella.
\newblock Batvision: Learning to see 3d spatial layout with two ears.
\newblock In \emph{2020 IEEE International Conference on Robotics and
  Automation (ICRA)}, pages 1581--1587. IEEE, 2020.

\bibitem[Gao et~al.(2020)Gao, Chen, Al-Halah, Schissler, and
  Grauman]{gao2020visualechoes}
R.~Gao, C.~Chen, Z.~Al-Halah, C.~Schissler, and K.~Grauman.
\newblock Visualechoes: Spatial image representation learning through
  echolocation.
\newblock In \emph{European Conference on Computer Vision}, pages 658--676.
  Springer, 2020.

\bibitem[Chen et~al.(2020)Chen, Jain, Schissler, Gari, Al-Halah, Ithapu,
  Robinson, and Grauman]{chen2020soundspaces}
C.~Chen, U.~Jain, C.~Schissler, S.~V.~A. Gari, Z.~Al-Halah, V.~K. Ithapu,
  P.~Robinson, and K.~Grauman.
\newblock Soundspaces: Audio-visual navigation in 3d environments.
\newblock In \emph{Proceedings of the European Conference on Computer Vision
  (ECCV)}. Springer, 2020.

\bibitem[de~Sa(1994)]{de1994learning}
V.~R. de~Sa.
\newblock Learning classification with unlabeled data.
\newblock In \emph{Advances in neural information processing systems}, pages
  112--119. Citeseer, 1994.

\bibitem[Ngiam et~al.(2011)Ngiam, Khosla, Kim, Nam, Lee, and
  Ng]{ngiam2011multimodal}
J.~Ngiam, A.~Khosla, M.~Kim, J.~Nam, H.~Lee, and A.~Y. Ng.
\newblock Multimodal deep learning.
\newblock In \emph{ICML}, 2011.

\bibitem[Owens et~al.(2016)Owens, Isola, McDermott, Torralba, Adelson, and
  Freeman]{owens2016visually}
A.~Owens, P.~Isola, J.~McDermott, A.~Torralba, E.~H. Adelson, and W.~T.
  Freeman.
\newblock Visually indicated sounds.
\newblock In \emph{CVPR}, 2016.

\bibitem[Arandjelovic and Zisserman(2017)]{arandjelovic2017look}
R.~Arandjelovic and A.~Zisserman.
\newblock Look, listen and learn.
\newblock In \emph{Proceedings of the IEEE International Conference on Computer
  Vision}, pages 609--617, 2017.

\bibitem[Owens and Efros(2018)]{owens2018learning}
A.~Owens and A.~A. Efros.
\newblock Audio-visual scene analysis with self-supervised multisensory
  features.
\newblock In \emph{ECCV}, 2018.

\bibitem[Korbar et~al.(2018)Korbar, Tran, and Torresani]{korbar2018cooperative}
B.~Korbar, D.~Tran, and L.~Torresani.
\newblock Cooperative learning of audio and video models from self-supervised
  synchronization.
\newblock \emph{arXiv preprint arXiv:1807.00230}, 2018.

\bibitem[Xiao et~al.(2020)Xiao, Lee, Grauman, Malik, and
  Feichtenhofer]{xiao2020audiovisual}
F.~Xiao, Y.~J. Lee, K.~Grauman, J.~Malik, and C.~Feichtenhofer.
\newblock Audiovisual slowfast networks for video recognition.
\newblock \emph{arXiv preprint arXiv:2001.08740}, 2020.

\bibitem[Morgado et~al.(2021)Morgado, Vasconcelos, and Misra]{morgado2021audio}
P.~Morgado, N.~Vasconcelos, and I.~Misra.
\newblock Audio-visual instance discrimination with cross-modal agreement.
\newblock In \emph{Proceedings of the IEEE/CVF Conference on Computer Vision
  and Pattern Recognition}, pages 12475--12486, 2021.

\bibitem[Asano et~al.(2020)Asano, Patrick, Rupprecht, and
  Vedaldi]{asano2020labelling}
Y.~M. Asano, M.~Patrick, C.~Rupprecht, and A.~Vedaldi.
\newblock Labelling unlabelled videos from scratch with multi-modal
  self-supervision.
\newblock \emph{arXiv preprint arXiv:2006.13662}, 2020.

\bibitem[Aytar et~al.(2016)Aytar, Vondrick, and Torralba]{aytar2016soundnet}
Y.~Aytar, C.~Vondrick, and A.~Torralba.
\newblock Soundnet: Learning sound representations from unlabeled video.
\newblock In \emph{Advances in neural information processing systems}, pages
  892--900, 2016.

\bibitem[Senocak et~al.(2018)Senocak, Oh, Kim, Yang, and
  So~Kweon]{senocak2018learning}
A.~Senocak, T.-H. Oh, J.~Kim, M.-H. Yang, and I.~So~Kweon.
\newblock Learning to localize sound source in visual scenes.
\newblock In \emph{Proceedings of the IEEE Conference on Computer Vision and
  Pattern Recognition}, pages 4358--4366, 2018.

\bibitem[Arandjelovic and Zisserman(2018)]{arandjelovic2018objects}
R.~Arandjelovic and A.~Zisserman.
\newblock Objects that sound.
\newblock In \emph{Proceedings of the European Conference on Computer Vision
  (ECCV)}, pages 435--451, 2018.

\bibitem[Harwath et~al.(2018)Harwath, Recasens, Sur{\'\i}s, Chuang, Torralba,
  and Glass]{harwath2018jointly}
D.~Harwath, A.~Recasens, D.~Sur{\'\i}s, G.~Chuang, A.~Torralba, and J.~Glass.
\newblock Jointly discovering visual objects and spoken words from raw sensory
  input.
\newblock In \emph{Proceedings of the European conference on computer vision
  (ECCV)}, pages 649--665, 2018.

\bibitem[Tian et~al.(2018)Tian, Shi, Li, Duan, and Xu]{tian2018audio}
Y.~Tian, J.~Shi, B.~Li, Z.~Duan, and C.~Xu.
\newblock Audio-visual event localization in unconstrained videos.
\newblock In \emph{Proceedings of the European Conference on Computer Vision
  (ECCV)}, pages 247--263, 2018.

\bibitem[Afouras et~al.(2020)Afouras, Owens, Chung, and
  Zisserman]{afouras2020self}
T.~Afouras, A.~Owens, J.~S. Chung, and A.~Zisserman.
\newblock Self-supervised learning of audio-visual objects from video.
\newblock \emph{arXiv preprint arXiv:2008.04237}, 2020.

\bibitem[Qian et~al.(2020)Qian, Hu, Dinkel, Wu, Xu, and Lin]{qian2020multiple}
R.~Qian, D.~Hu, H.~Dinkel, M.~Wu, N.~Xu, and W.~Lin.
\newblock Multiple sound sources localization from coarse to fine.
\newblock \emph{arXiv preprint arXiv:2007.06355}, 2020.

\bibitem[Hu et~al.(2020)Hu, Qian, Jiang, Tan, Wen, Ding, Lin, and
  Dou]{hu2020discriminative}
D.~Hu, R.~Qian, M.~Jiang, X.~Tan, S.~Wen, E.~Ding, W.~Lin, and D.~Dou.
\newblock Discriminative sounding objects localization via self-supervised
  audiovisual matching.
\newblock \emph{Advances in Neural Information Processing Systems}, 33, 2020.

\bibitem[Chen et~al.(2021)Chen, Xie, Afouras, Nagrani, Vedaldi, and
  Zisserman]{chen2021localizing}
H.~Chen, W.~Xie, T.~Afouras, A.~Nagrani, A.~Vedaldi, and A.~Zisserman.
\newblock Localizing visual sounds the hard way.
\newblock \emph{arXiv preprint arXiv:2104.02691}, 2021.

\bibitem[Chung and Zisserman(2016)]{chung2016out}
J.~S. Chung and A.~Zisserman.
\newblock Out of time: automated lip sync in the wild.
\newblock In \emph{Asian conference on computer vision}, pages 251--263.
  Springer, 2016.

\bibitem[Chung et~al.(2019)Chung, Chung, and Kang]{chung2019perfect}
S.-W. Chung, J.~S. Chung, and H.-G. Kang.
\newblock Perfect match: Improved cross-modal embeddings for audio-visual
  synchronisation.
\newblock In \emph{ICASSP 2019-2019 IEEE International Conference on Acoustics,
  Speech and Signal Processing (ICASSP)}, pages 3965--3969. IEEE, 2019.

\bibitem[Roth et~al.(2020)Roth, Chaudhuri, Klejch, Marvin, Gallagher, Kaver,
  Ramaswamy, Stopczynski, Schmid, Xi, et~al.]{roth2020ava}
J.~Roth, S.~Chaudhuri, O.~Klejch, R.~Marvin, A.~Gallagher, L.~Kaver,
  S.~Ramaswamy, A.~Stopczynski, C.~Schmid, Z.~Xi, et~al.
\newblock Ava active speaker: An audio-visual dataset for active speaker
  detection.
\newblock In \emph{ICASSP 2020-2020 IEEE International Conference on Acoustics,
  Speech and Signal Processing (ICASSP)}, pages 4492--4496. IEEE, 2020.

\bibitem[Gao et~al.(2018)Gao, Feris, and Grauman]{gao2018learning}
R.~Gao, R.~Feris, and K.~Grauman.
\newblock Learning to separate object sounds by watching unlabeled video.
\newblock In \emph{ECCV}, 2018.

\bibitem[Zhao et~al.(2018)Zhao, Gan, Rouditchenko, Vondrick, McDermott, and
  Torralba]{zhao2018sound}
H.~Zhao, C.~Gan, A.~Rouditchenko, C.~Vondrick, J.~McDermott, and A.~Torralba.
\newblock The sound of pixels.
\newblock In \emph{ECCV}, 2018.

\bibitem[Zhao et~al.(2019)Zhao, Gan, Ma, and Torralba]{zhao2019sound}
H.~Zhao, C.~Gan, W.-C. Ma, and A.~Torralba.
\newblock The sound of motions.
\newblock In \emph{Proceedings of the IEEE/CVF International Conference on
  Computer Vision}, pages 1735--1744, 2019.

\bibitem[Gao and Grauman(2021)]{gao2021visualvoice}
R.~Gao and K.~Grauman.
\newblock Visualvoice: Audio-visual speech separation with cross-modal
  consistency.
\newblock \emph{arXiv preprint arXiv:2101.03149}, 2021.

\bibitem[Hoiem et~al.(2007)Hoiem, Efros, and Hebert]{hoiem2007recovering}
D.~Hoiem, A.~A. Efros, and M.~Hebert.
\newblock Recovering surface layout from an image.
\newblock \emph{International Journal of Computer Vision}, 75\penalty0
  (1):\penalty0 151--172, 2007.

\bibitem[Saxena et~al.(2008)Saxena, Sun, and Ng]{saxena2008make3d}
A.~Saxena, M.~Sun, and A.~Y. Ng.
\newblock Make3d: Learning 3d scene structure from a single still image.
\newblock \emph{IEEE transactions on pattern analysis and machine
  intelligence}, 31\penalty0 (5):\penalty0 824--840, 2008.

\bibitem[Eigen and Fergus(2015)]{eigen2015predicting}
D.~Eigen and R.~Fergus.
\newblock Predicting depth, surface normals and semantic labels with a common
  multi-scale convolutional architecture.
\newblock In \emph{ICCV}, 2015.

\bibitem[Chen et~al.(2016)Chen, Fu, Yang, and Deng]{chen2016single}
W.~Chen, Z.~Fu, D.~Yang, and J.~Deng.
\newblock Single-image depth perception in the wild.
\newblock In \emph{Advances in neural information processing systems}, pages
  730--738, 2016.

\bibitem[Chen et~al.(2019)Chen, Qian, and Deng]{chen2019learning}
W.~Chen, S.~Qian, and J.~Deng.
\newblock Learning single-image depth from videos using quality assessment
  networks.
\newblock In \emph{CVPR}, 2019.

\bibitem[Li and Snavely(2018)]{li2018megadepth}
Z.~Li and N.~Snavely.
\newblock Megadepth: Learning single-view depth prediction from internet
  photos.
\newblock In \emph{Proceedings of the IEEE Conference on Computer Vision and
  Pattern Recognition}, pages 2041--2050, 2018.

\bibitem[Godard et~al.(2019)Godard, Mac~Aodha, Firman, and
  Brostow]{godard2019digging}
C.~Godard, O.~Mac~Aodha, M.~Firman, and G.~J. Brostow.
\newblock Digging into self-supervised monocular depth estimation.
\newblock In \emph{Proceedings of the IEEE/CVF International Conference on
  Computer Vision}, pages 3828--3838, 2019.

\bibitem[Tulsiani et~al.(2018)Tulsiani, Gupta, Fouhey, Efros, and
  Malik]{tulsiani2018factoring}
S.~Tulsiani, S.~Gupta, D.~F. Fouhey, A.~A. Efros, and J.~Malik.
\newblock Factoring shape, pose, and layout from the 2d image of a 3d scene.
\newblock In \emph{CVPR}, 2018.

\bibitem[Gkioxari et~al.(2019)Gkioxari, Malik, and Johnson]{gkioxari2019mesh}
G.~Gkioxari, J.~Malik, and J.~Johnson.
\newblock Mesh r-cnn.
\newblock In \emph{ICCV}, 2019.

\bibitem[Girdhar et~al.(2016)Girdhar, Fouhey, Rodriguez, and
  Gupta]{girdhar2016learning}
R.~Girdhar, D.~F. Fouhey, M.~Rodriguez, and A.~Gupta.
\newblock Learning a predictable and generative vector representation for
  objects.
\newblock In \emph{ECCV}. Springer, 2016.

\bibitem[Yin et~al.(2020)Yin, Zhang, Wang, Niklaus, Mai, Chen, and
  Shen]{yin2020learning}
W.~Yin, J.~Zhang, O.~Wang, S.~Niklaus, L.~Mai, S.~Chen, and C.~Shen.
\newblock Learning to recover 3d scene shape from a single image.
\newblock \emph{arXiv preprint arXiv:2012.09365}, 2020.

\bibitem[Hartley and Zisserman(2004)]{hartley2004mvg}
R.~I. Hartley and A.~Zisserman.
\newblock \emph{Multiple View Geometry in Computer Vision}.
\newblock Cambridge University Press, ISBN: 0521540518, second edition, 2004.

\bibitem[Hartley(1997)]{hartley1997defense}
R.~I. Hartley.
\newblock In defense of the eight-point algorithm.
\newblock \emph{IEEE Transactions on pattern analysis and machine
  intelligence}, 19\penalty0 (6):\penalty0 580--593, 1997.

\bibitem[Wu et~al.(2008)Wu, Clipp, Li, Frahm, and Pollefeys]{wu20083d}
C.~Wu, B.~Clipp, X.~Li, J.-M. Frahm, and M.~Pollefeys.
\newblock 3d model matching with viewpoint-invariant patches (vip).
\newblock In \emph{2008 IEEE Conference on Computer Vision and Pattern
  Recognition}, pages 1--8. IEEE, 2008.

\bibitem[Qian et~al.(2020)Qian, Jin, and Fouhey]{qian2020associative3d}
S.~Qian, L.~Jin, and D.~F. Fouhey.
\newblock Associative3d: Volumetric reconstruction from sparse views.
\newblock In \emph{European Conference on Computer Vision}, pages 140--157.
  Springer, 2020.

\bibitem[Cai et~al.(2021)Cai, Hariharan, Snavely, and
  Averbuch-Elor]{cai2021extreme}
R.~Cai, B.~Hariharan, N.~Snavely, and H.~Averbuch-Elor.
\newblock Extreme rotation estimation using dense correlation volumes.
\newblock \emph{arXiv preprint arXiv:2104.13530}, 2021.

\bibitem[Zhu et~al.(2017)Zhu, Mottaghi, Kolve, Lim, Gupta, Fei-Fei, and
  Farhadi]{zhu2017target}
Y.~Zhu, R.~Mottaghi, E.~Kolve, J.~J. Lim, A.~Gupta, L.~Fei-Fei, and A.~Farhadi.
\newblock Target-driven visual navigation in indoor scenes using deep
  reinforcement learning.
\newblock In \emph{2017 IEEE international conference on robotics and
  automation (ICRA)}, pages 3357--3364. IEEE, 2017.

\bibitem[Mirowski et~al.(2016)Mirowski, Pascanu, Viola, Soyer, Ballard, Banino,
  Denil, Goroshin, Sifre, Kavukcuoglu, et~al.]{mirowski2016learning}
P.~Mirowski, R.~Pascanu, F.~Viola, H.~Soyer, A.~J. Ballard, A.~Banino,
  M.~Denil, R.~Goroshin, L.~Sifre, K.~Kavukcuoglu, et~al.
\newblock Learning to navigate in complex environments.
\newblock \emph{arXiv preprint arXiv:1611.03673}, 2016.

\bibitem[Savva et~al.(2019)Savva, Kadian, Maksymets, Zhao, Wijmans, Jain,
  Straub, Liu, Koltun, Malik, et~al.]{savva2019habitat}
M.~Savva, A.~Kadian, O.~Maksymets, Y.~Zhao, E.~Wijmans, B.~Jain, J.~Straub,
  J.~Liu, V.~Koltun, J.~Malik, et~al.
\newblock Habitat: A platform for embodied ai research.
\newblock In \emph{Proceedings of the IEEE/CVF International Conference on
  Computer Vision}, pages 9339--9347, 2019.

\bibitem[Wijmans et~al.(2019)Wijmans, Kadian, Morcos, Lee, Essa, Parikh, Savva,
  and Batra]{wijmans2019dd}
E.~Wijmans, A.~Kadian, A.~Morcos, S.~Lee, I.~Essa, D.~Parikh, M.~Savva, and
  D.~Batra.
\newblock Dd-ppo: Learning near-perfect pointgoal navigators from 2.5 billion
  frames.
\newblock \emph{arXiv preprint arXiv:1911.00357}, 2019.

\bibitem[Rascon and Meza(2017)]{rascon2017localization}
C.~Rascon and I.~Meza.
\newblock Localization of sound sources in robotics: A review.
\newblock \emph{Robotics and Autonomous Systems}, 96:\penalty0 184--210, 2017.

\bibitem[An et~al.(2018)An, Son, Manocha, and Yoon]{an2018reflection}
I.~An, M.~Son, D.~Manocha, and S.-E. Yoon.
\newblock Reflection-aware sound source localization.
\newblock In \emph{2018 IEEE International Conference on Robotics and
  Automation (ICRA)}, pages 66--73. IEEE, 2018.

\bibitem[Gan et~al.(2019)Gan, Zhang, Wu, Gong, and Tenenbaum]{gan2019look}
C.~Gan, Y.~Zhang, J.~Wu, B.~Gong, and J.~B. Tenenbaum.
\newblock Look, listen, and act: Towards audio-visual embodied navigation.
\newblock \emph{arXiv preprint arXiv:1912.11684}, 2019.

\bibitem[Chen et~al.(2020{\natexlab{a}})Chen, Majumder, Al-Halah, Gao,
  Ramakrishnan, and Grauman]{chen2020learning}
C.~Chen, S.~Majumder, Z.~Al-Halah, R.~Gao, S.~K. Ramakrishnan, and K.~Grauman.
\newblock Learning to set waypoints for audio-visual navigation.
\newblock \emph{arXiv preprint arXiv:2008.09622}, 1\penalty0 (2):\penalty0 6,
  2020{\natexlab{a}}.

\bibitem[Chen et~al.(2020{\natexlab{b}})Chen, Al-Halah, and
  Grauman]{chen2020semantic}
C.~Chen, Z.~Al-Halah, and K.~Grauman.
\newblock Semantic audio-visual navigation.
\newblock \emph{arXiv preprint arXiv:2012.11583}, 2020{\natexlab{b}}.

\bibitem[Palmer(1999)]{palmer1999vision}
S.~E. Palmer.
\newblock \emph{Vision science: Photons to phenomenology}.
\newblock MIT press, 1999.

\bibitem[Gandhi et~al.(2017)Gandhi, Pinto, and Gupta]{gandhi2017learning}
D.~Gandhi, L.~Pinto, and A.~Gupta.
\newblock Learning to fly by crashing.
\newblock In \emph{2017 IEEE/RSJ International Conference on Intelligent Robots
  and Systems (IROS)}, pages 3948--3955. IEEE, 2017.

\bibitem[Hershey et~al.(2017)Hershey, Chaudhuri, Ellis, Gemmeke, Jansen, Moore,
  Plakal, Platt, Saurous, Seybold, et~al.]{hershey2017cnn}
S.~Hershey, S.~Chaudhuri, D.~P. Ellis, J.~F. Gemmeke, A.~Jansen, R.~C. Moore,
  M.~Plakal, D.~Platt, R.~A. Saurous, B.~Seybold, et~al.
\newblock Cnn architectures for large-scale audio classification.
\newblock In \emph{2017 ieee international conference on acoustics, speech and
  signal processing (icassp)}, pages 131--135. IEEE, 2017.

\bibitem[He et~al.(2016)He, Zhang, Ren, and Sun]{he2016deep}
K.~He, X.~Zhang, S.~Ren, and J.~Sun.
\newblock Deep residual learning for image recognition.
\newblock In \emph{Proceedings of the IEEE conference on computer vision and
  pattern recognition}, pages 770--778, 2016.

\bibitem[Deng et~al.(2009)Deng, Dong, Socher, Li, Li, and
  Fei-Fei]{deng2009imagenet}
J.~Deng, W.~Dong, R.~Socher, L.-J. Li, K.~Li, and L.~Fei-Fei.
\newblock Imagenet: A large-scale hierarchical image database.
\newblock In \emph{2009 IEEE conference on computer vision and pattern
  recognition}, pages 248--255. Ieee, 2009.

\bibitem[Sutskever et~al.(2013)Sutskever, Martens, Dahl, and
  Hinton]{sutskever2013importance}
I.~Sutskever, J.~Martens, G.~Dahl, and G.~Hinton.
\newblock On the importance of initialization and momentum in deep learning.
\newblock In \emph{International conference on machine learning}, pages
  1139--1147. PMLR, 2013.

\bibitem[Kingma and Ba(2015)]{kingma2014adam}
D.~P. Kingma and J.~Ba.
\newblock Adam: A method for stochastic optimization.
\newblock 2015.

\bibitem[Chen et~al.(2020)Chen, Xie, Vedaldi, and Zisserman]{chen2020vggsound}
H.~Chen, W.~Xie, A.~Vedaldi, and A.~Zisserman.
\newblock Vggsound: A large-scale audio-visual dataset.
\newblock In \emph{ICASSP 2020-2020 IEEE International Conference on Acoustics,
  Speech and Signal Processing (ICASSP)}, pages 721--725. IEEE, 2020.

\bibitem[Fonseca et~al.(2017)Fonseca, Pons~Puig, Favory, Font~Corbera,
  Bogdanov, Ferraro, Oramas, Porter, and Serra]{fonseca2017freesound}
E.~Fonseca, J.~Pons~Puig, X.~Favory, F.~Font~Corbera, D.~Bogdanov, A.~Ferraro,
  S.~Oramas, A.~Porter, and X.~Serra.
\newblock Freesound datasets: a platform for the creation of open audio
  datasets.
\newblock In \emph{Hu X, Cunningham SJ, Turnbull D, Duan Z, editors.
  Proceedings of the 18th ISMIR Conference; 2017 oct 23-27; Suzhou,
  China.[Canada]: International Society for Music Information Retrieval; 2017.
  p. 486-93.} International Society for Music Information Retrieval (ISMIR),
  2017.

\bibitem[Vaswani et~al.(2017)Vaswani, Shazeer, Parmar, Uszkoreit, Jones, Gomez,
  Kaiser, and Polosukhin]{vaswani2017attention}
A.~Vaswani, N.~Shazeer, N.~Parmar, J.~Uszkoreit, L.~Jones, A.~N. Gomez,
  {\L}.~Kaiser, and I.~Polosukhin.
\newblock Attention is all you need.
\newblock In \emph{Advances in neural information processing systems}, pages
  5998--6008, 2017.

\end{thebibliography}

\clearpage
\renewcommand{\thesection}{A.\arabic{section}}
\setcounter{section}{0}

\section{Dataset Examples}
We show video samples of \emph{\audioum} recordings and \emph{\audiotd} recordings along with audio in our project webpage, check \hyperlink{https://ificl.github.io/structure-from-silence/}{here} for details. 

\section{Additional robot navigation experiments}
\label{appendix:robot}
We perform additional robotic navigation experiments. For these, we use a robot equipped with to avoid the need for the robot to rotate to sense both sides of the scene (the model is otherwise unchanged). 

\begin{figure*}[!tb]
    \vspace{-3.0em}
    \centering
     \includegraphics[width=\textwidth]{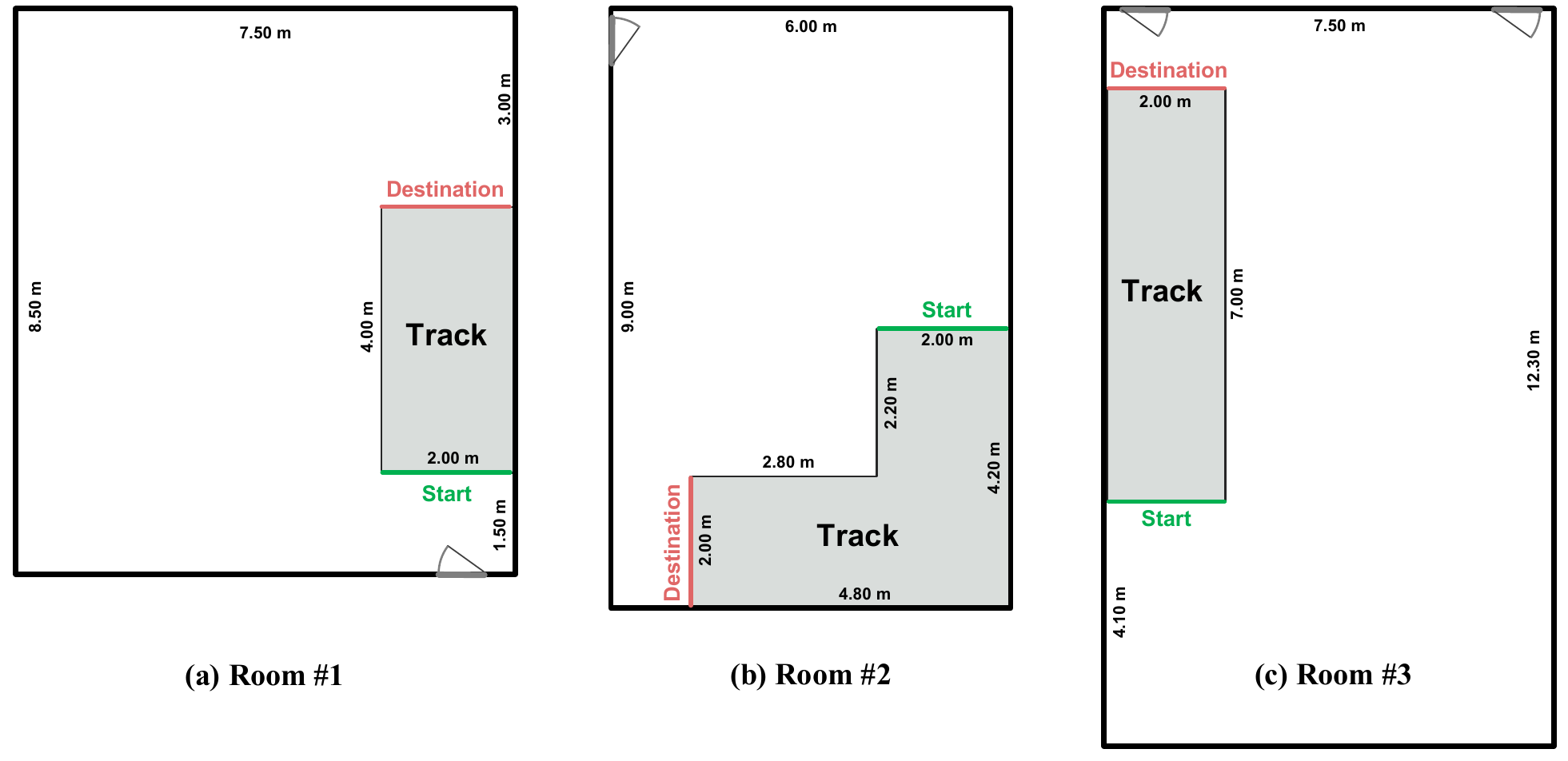}
    \caption{Classroom floor plans and corresponding track settings. } 
    \label{fig:suppfloorplan}
\end{figure*}

\mypar{Experiment setting.} We evaluate our model in 3 additional rooms. Their floor plans and track designs are shown in Fig.~\ref{fig:suppfloorplan}. Similarly, the robot starts from one of several unknown locations (40cm, 80cm, or 120cm from the wall) and orientations ($30^{\circ}$ left or right, or facing forward). For each room, we repeat the experiment 9 times and measure the longest distance along the track it attained before colliding with a boundary. For the \textquotesingle L\textquotesingle-shaped track in Room \#2, we determine the distance by projecting the final position to the center line of the track.  All rooms (4 in total) are located in distinct buildings, and are not included in the training set. 

\mypar{Algorithm.} The detailed algorithm to control the robot is shown in Algorithm~\ref{alg:robot-policy}. When computing the near-wall probability $P(L)$ and $P(R)$, we average the predictions of 20 clips of 3s audio to make the prediction more accurate robust. For each room, we manually adjusted the threshold $p$, which accounts for per-room biases in the classifier's output.
\begin{wrapfigure}[19]{R}{0.6\textwidth}
\vspace{-5mm}
\hspace{-2.5mm}
    \input{floats/fig_supp_robot_res}
\end{wrapfigure}
\mypar{Experimental results.} As before, we evaluate using the ``straight line'' policy. We measure the performance by computing the percentage of the completed track, and take the average of all trials, as well as the average of the trials starting with the same orientation, as shown in Fig.~\ref{fig:supp_navig}.  It can be seen that the ``straight line'' policy performs well when reaching the goal does not require turning, as expected. However, it performs significantly worse than our model when the robot starts at an angle that does not follow the wall. In contrast, the distance our model traverses is more consistent between starting orientations, indicating that the model more successfully guides the robot to navigate along the wall. When looking at the average performance with different starting orientation on individual rooms, we also find that the ``straight line'' policy only outperforms our method when starting facing forward on straight tracks, and performs worse than ours with any other starting orientation or on the \textquotesingle L\textquotesingle-shaped track in Room \#2. 

\mypar{Vision model.} While, in principle, vision-based navigation is very effective (and is capable of significantly outperforming audio-based navigation), we found in initial experiments that our vision-based model did not generalize well to novel viewpoints, perhaps because the RGB-D training only contained wall-facing viewpoints. To address this, we collected 12K RGB-D images as training data from the same scenes, with additional random viewing directions, and finetune our visual model on it. We then used a slightly modified algorithm (Algorithm~\ref{alg:robot-policy-vision}). To keep the control algorithm as close as possible to the controller for audio model, we made a small modification to Algorithm~\ref{alg:robot-policy}. We estimated the turning angle based on the prediction of near-wall side (which can be obtained by comparing left-side score and right-side score at the initial step). The reason for this is to avoid the noise from the prediction of far-wall side, considering the threshold $p$ is usually very high (we set it as 0.97 in the experiment).

\mypar{Video demo.} We provided demo video in our project webpage. Please check the video for qualitative results of robot navigation.  

\clearpage

\section{Obstacle detection on \audioum-dense recordings}
\label{appendix:static-dense}
\begin{wraptable}{r}{0.4\textwidth}
\vspace{-1.2em}
  \caption{{\bf Obstacle detection} for \emph{\audioum-dense} recordings. Here, \textbf{S} denotes \textit{static} and \textbf{M} means \textit{motion}.}
  \label{tab:obstacle_dense}
  \centering
  \resizebox{\linewidth}{!}
  {
  \begin{tabular}{lccc} 
    \toprule
    Model     & Task & AP(\%) & Acc(\%)\\
    \midrule
    Chance & S & 46.4 &  50.0\\
    Chance  & M & 52.4 &  49.8\\
    Ours  & S & \textbf{96.4} &  \textbf{83.9}\\
    Ours  & M & \textbf{98.3} &  \textbf{94.2}\\
    Ours-static & S & 93.9 &  84.6\\
    
    \bottomrule
  \end{tabular}
  }

\vspace{-1.0em}
\end{wraptable} 

We evaluated the obstacle detection task for \emph{\audioum-dense} recordings by training a model on 3 rooms and testing on an unseen room with the same method as the main paper. 
To evaluate whether the success rate for wall perception increases when participants have self-motion, we simulate the motion of the recorder by concatenating two audio with the same angle from adjacent grids, and predict whether the agent is moving towards or far away from the obstacles. We report the performance of the model~({\bf Ours-static}) trained on \emph{\audioum} recordings for comparison as well.

As shown in Table~\ref{tab:obstacle_dense}, the average precision above 90\%  in \emph{\audioum-dense} recordings with both settings. 
Comparing experimental results of with or without motion, we can find the network predicts more accurately with the simulated motion.  
We also evaluate our \emph{\audioum-dense} model on \emph{\audioum} recordings, finding that it could obtain 62.8\% AP and 60.9\% accuracy (on par with the equivalent model trained on the \emph{\audioum} dataset).
A video demo is available in our project webpage.

\section{Indoor localization on \audioum-dense recordings} 
We also asked whether ambient sounds can convey the absolute position in rooms (rather than distance to walls), when trained and tested on similar (or the same) room. We performed a localization experiment for \emph{\audioum-dense} recordings. We use the same network as the obstacle detection task and replace the last linear layer with a $N$-class fully-connected layer, where $N$ is the number of grid cells in the room. We formulate this as a multi-way classification task and predict which grid the given audio comes from. We train the model with the cross entropy loss and evaluate it with top-1, top-5, and average distance (\ie, the Euclidean distance between our predicted and target grid position).

\vspace{0.1em}
\textbf{Grid classification.} We ask to what extent ambient sound can allow a model to predict which grid cell within the {\em same room} the model was trained on. For the train/test split, we randomly select audio samples with three angles in each grid cell as training, then test on sound with the remaining angle (i.e. the test samples are unseen, but the model has been trained with other examples from the same room).
The results are shown in Table~\ref{tab:indoor_localization}. The model's accuracy is significantly better than chance. 
As expected, we see that our absolute-position model does not generalize well to other rooms, which may be due to both ambiguities in the coordinate system and fundamental ambiguities in the prediction problem. %

\noindent
\textbf{Generalization over time.} 
Next, we evaluated our model's ability to predict the absolute position in the room using the audio samples recorded at different times to see how well the model could generalize over time. In this experiment, training and test sets are from the same room but different times.
The time intervals for Room \#4, \#5 and \#6 are 1 week, 1 week, and 1 hour, respectively.

We show the experimental results in Table~\ref{tab:indoor_localization}.
It can be seen that the performance is lower when audio is recorded at different times, confirming the assumption that the network can indeed use the shortcut of memorizing transient background sounds. However, the performance is still significantly better than random chance for recordings that were taken a week apart. %
\begin{table}[t]
\vspace{-2.0em}
  \caption{{\bf Indoor Localization} for \emph{\audioum-dense} recordings. }
  \label{tab:indoor_localization}
  \centering
   \resizebox{\linewidth}{!}
{
  \begin{tabular}{lccc@{\hskip20pt}ccc@{\hskip20pt}ccc}%
    \toprule
       & \multicolumn{3}{c@{\hskip20pt}}{Room 4} & \multicolumn{3}{c@{\hskip20pt}}{Room 5} & \multicolumn{3}{c}{Room 6} \\
    EXP.  & Random & Non time-shift  & Time-shift & Random & Non time-shift  & Time-shift & Random & Non time-shift  & Time-shift \\
    \midrule
    Top-1~(\%)   & 4.00  & 32.3  & 18.3  & 2.86  & 29.5   & 18.8 & 4.00  & 34.3   & 22.8   \\
    Top-5~(\%)   & 20.0  & 77.7  & 50.3  & 14.5  & 70.8   & 51.8  & 20.0  & 74.3   & 55.6 \\
    Avg. Distance     & 3.17  & 1.23  & 1.71  & 3.59  & 1.98   & 2.17  & 3.17  & 1.31   & 1.87  \\
    \bottomrule
  \end{tabular}
  }
\end{table}

\section{Relative depth order on {\audiotd} recordings}
We design experiments to investigate the accuracy of relative depth estimation model as the magnitude of the depth difference changes. We paired audio in the test set with 5 different distance ranges, and re-evaluated our model (without re-training). We used the variations of our networks that are initialized randomly. %
The results are shown in Figure~\ref{fig:pairdis}. We see that prediction accuracy increases as the relative depth increases, and that the models are more confident when the distances apart are at least 0.2 meters. 

\section{Generalization between \textit{motion} and \textit{static} recordings}

\begin{wraptable}{r}{0.4\textwidth}
\vspace{-1.2em}
  \caption{{\bf Generalization between \textit{motion} and \textit{static} recordings }on obstacle detection and depth order task. }
  \label{tab:genaralization_static_motion}
  \centering
   \resizebox{\linewidth}{!}
{
  \begin{tabular}{lcc}
    \toprule
     & AP(\%) & Acc(\%) \\
    \midrule
    Obstacle detection   & 61.3 & 51.9   \\
    Depth order   & 78.8 & 70.2 \\
    \bottomrule
  \end{tabular}
  }
\vspace{-3.0em}
\end{wraptable} 
We give the results of the model trained on motion recordings (recorded in hallways) and tested on static recording (recorded in the classrooms). The experiment results are shown in Table~\ref{tab:genaralization_static_motion} . Our model can still perform well above random chance despite evaluating on a different domain.

\begin{figure*}[b]
\vspace{-1.0em}
    \centering
    \begin{minipage}[c][1\width]{0.48\textwidth}
        \input{floats/fig_pairdis.tex}
    \end{minipage}
    \hfill
    \begin{minipage}[c][1\width]{0.48\textwidth}
        \input{floats/fig_nonambient_curve}
    \end{minipage}
    
\end{figure*}

\section{Effects of Non-ambient sounds}
\label{appendix:nonambient}
To investigate how non-ambient sounds from environment could affect our prediction, we mix {\audioum} recordings with sounds from FreeSound~\citep{fonseca2017freesound} to simulate other sound sources in the scenes. By using synthetic mixtures, we prevent our model from exploiting other cues, \eg echolocation. We use 200 types of sound events at training and test time, \eg, human sounds, animal, and music. %
We test our models with (or without) re-training on the mixed sounds. The results are shown in Fig.~\ref{fig:nonambient}. We can see that the performance drops as the volume of distracting sound increases, while still outperforming chance. Re-training the model with mixed sounds improves performance and makes the model significantly more robust.

\begin{figure*}[t]
\vspace{-3.0em}
    \centering
    \begin{minipage}[c][1\width]{0.39\textwidth}
        \input{floats/fig_CADmodel.tex}
    \end{minipage}
    \hfill
    \begin{minipage}[c][1\width]{0.59\textwidth}
        \input{floats/fig_avorder}
    \end{minipage}
\vspace{-0.5em}
\end{figure*}

\section{Conditional absolute depth estimation}
\label{appendix:conditional_ade}
\mypar{Implementation.} We use a Siamese network from relative tasks to build conditional absolute depth~(\textbf{CAD}) model. The model takes three inputs: audio $s_1$, reference audio $s_2$, and ground-truth depth $x_2$. We first extract audio features with VGGish backbone and calculate the depth embedding following~\citep{vaswani2017attention}. We add  the depth embedding and reference audio feature and fuse features via concatenation prior to the multi-layer perceptron~(\ie, FC-ReLU-FC-ReLU-FC layers). During training, the reference sounds are randomly selected from the same scene, while those are fixed for each sample in the test set ~(Fig.~\ref{fig:cad_model}).

\section{Self-supervised audio-visual learning}
\label{appendix:self-supervision}
\mypar{Implementation.} 
We approach the AV-Order model by building two Siamese networks for both audio and visual modalities. As shown in Fig.~\ref{fig:avorder_model}, for each modality, we first extract features from inputs and concatenate them prior to the multi-layer perceptron. Then we concatenate two 128-d feature vectors from image and audio branches, and forward it to an MLP~(\ie, FC-ReLU-FC layers, and all the MLP blocks are the same) for the logits.

\mypar{AudioSet features.} 
The low performance of the AudioSet features in Table~\ref{tab:avco_binary}, \ref{tab:avco_multi} could be due to
differences in the types of sounds considered, as well as the preprocessing in each method. The preprocessing procedure used in the AudioSet model (which is standard on the dataset) throws away all but the 125Hz-7500Hz frequency range, which our analysis (Section \ref{sec:audio_analysis}) suggests is important. To test this, we trained a random feature baseline with (and without) this preprocessing, and see a large difference in their performances~(Table~\ref{tab:avco_binary}, \ref{tab:avco_multi}).

\mypar{Few-shot learning.}  
To help understand how audio-visual self-supervision improves fine-tuned models, we measure the performance using various numbers of labeled examples as training set~(5\%, 10\%, 20\%, 40\%, 100\%). The experiment results are shown in Figure~\ref{fig:fewshot}. We can see a large improvement from self-supervised initialization when there is more unlabeled data than labeled data, especially in few-shot training regimes. When the number of labeled examples approaches the number of unlabeled examples, the trained-from-scratch model catches up. This is expected, since the two datasets are exactly the same, and the labels provide strictly more information than the audio.

\begin{figure*}[b]
\vspace{-1.0em}
    \centering
    \begin{minipage}[c][1\width]{0.48\textwidth}
        \vspace{0.6em}
        \input{floats/fig_fewshot}
    \end{minipage}
    \hfill
    \begin{minipage}[c][1\width]{0.48\textwidth}
         \input{floats/fig_energy}
    \end{minipage}
    
\end{figure*}

\section{VGGSound-Instrument dataset}
\label{appendix:vggsound}
We sampled 37 classes of musical instruments with 32k video clips of 10s length from VGGSound dataset~\cite{chen2020vggsound}, and we call this subset VGGSound-Instruments. Video classes are listed as below.
We explore frequency energy distributions of the motion recordings and VGGSound-Instruments by computing the energy ratio, \ie the ratio between the magnitude in a given frequency and the sum of all frequency magnitudes, from all the samples. As shown in Fig.~\ref{fig:energy}, we can see that 70\% of the spectrogram energy is concentrated on low-frequencies (0 to 1000Hz), which may explain why our model achieves the best performance with low-frequency information in Fig.~\ref{fig:freq}. Furthermore, our ambient dataset contains relatively more low-frequency sounds than VGGSound-Instrument, suggesting the diversity of our dataset.

\vspace{-1em}
\hspace{-10mm}

\begin{table}[H]
\begin{center}
\hspace{-2em}
\resizebox{1.1\columnwidth}{!}{
\begin{tabular}{l l l l l}
playing accordion       &   playing acoustic guitar   &
playing banjo           &   playing bass drum   &
playing bass guitar     \\ 
playing bongo           &   playing cello           &   
playing clarinet        &   playing congas  &
playing cornet  \\
playing cymbal          &   playing djembe   &
playing double bass     &   playing drum kit   &
playing electric guitar \\ 
playing electronic organ &  playing erhu    &
playing flute   &   playing glockenspiel   &
playing guiro  \\
playing hammond organ   &   playing harp   &
playing harpsichord     &   playing mandolin   & 
playing marimba, xylophone   \\
playing piano   &  playing saxophone       &   
playing sitar   &   playing snare drum      &   
playing steel guitar, slide guitar \\
playing tabla           &   playing timbales  &
playing trumpet         &   playing ukulele &
playing vibraphone      \\  
playing violin, fiddle   &  playing zither          &    &  &       \\
\end{tabular}
}
\end{center}
\label{tab:category}
\end{table}

\vspace{8em}

\begin{figure*}[!h]
\hspace{-6mm}

    \centering
    \begin{minipage}[c][1\width]{0.48\textwidth}
        \input{floats/algo_robot}
    \end{minipage}
    \begin{minipage}[c][1\width]{0.48\textwidth}
        \input{floats/algo_robot_vision}
    \end{minipage}
\end{figure*}

\end{document}